# Graphene oxide films for ultra-flat optics and linear and nonlinear integrated photonic circuits

*Jiayang Wu[1], Linnan Jia[1], Yuning Zhang[1], Yang Qu[1], Baohua Jia [2,] * and David J. Moss [1,]*

[1]Optical Sciences Centre, Swinburne University of Technology, Hawthorn, VIC 3122, Australia

[2]Centre for Translational Atomaterials, Swinburne University of Technology, Hawthorn, VIC 3122, Australia

*E-mail: bjia@swin.edu.au, dmoss@swin.edu.au

**Abstract**

With superior optical properties, high flexibility in engineering its material properties, and strong capability for large-scale on-chip integration, graphene oxide (GO) is an attractive solution for on-chip integration of two-dimensional (2D) materials to implement functional integrated photonic devices capable of new features. Over the past decade, integrated GO photonics, representing an innovative merging of integrated photonic devices and thin GO films, has experienced significant development, leading to a surge in many applications covering almost every field of optical sciences. This paper reviews the recent advances in this emerging field, providing an overview of the optical properties of GO as well as methods for the on-chip integration of GO. The main achievements made in GO hybrid integrated photonic devices for diverse applications are summarized. The open challenges as well as the potential for future improvement are also discussed.

# 1. Introduction

The past decade has witnessed an enormous surge in activity in layered two-dimensional (2D) materials [1, 2]. Research on 2D materials was initially ignited by the ground-breaking work on graphene in 2004 [3] and has since expanded to other 2D materials such as graphene oxide (GO), transition metal dichalcogenides (TMDCs), hexagonal boron nitride (hBN), and black phosphorus (BP). The field has focused not only their distinctive electrical and chemical properties, but also their fascinating mechanical, thermal, and optical properties [2, 4]. In particular, in the field of optics, 2D materials exhibit many remarkable properties such as a broadband ultrafast optical response, large optical nonlinearities and strong material anisotropy. These have enabled many new photonic devices that are fundamentally different from those based on traditional bulk materials [2, 5-11].

Integrated platforms, particularly those compatible with the well-developed complementary metal-oxide-semiconductor (CMOS) fabrication technology, such as silicon, silicon nitride (SiN), and doped silica [12-14], have been widely exploited to implement integrated devices for many applications including telecommunications, IT services, displays, astronomy, sensing, and many others. Integrating 2D materials into these platforms offers the best of both worlds: not only does it benefit in terms of compact device footprint, high stability, and mass producibility, but it also enables new capabilities and significantly improves the device performance by exploiting the superior material properties of 2D materials.

The on-chip integration of 2D materials typically requires layer transfer processes [5, 7], where exfoliated or chemical vapour deposition grown 2D membranes are attached onto dielectric substrates (e.g., silicon and silica wafers). Despite its widespread implementation, the transfer approach for 2D materials is sophisticated, making it difficult to achieve large-area, highly uniform, and consistent coatings as well as the precise patterning needed for integrated devices [15]. This significantly limits the production scale for state-of-the-art integrated devices that incorporate 2D materials, hindering the practical application of 2D materials outside the laboratory.

The history of GO can in fact be traced back to 1859 [16]. It has been traditionally recognized as a precursor for the production of graphene-like materials and devices. In recent years it has attracted increasing interest in its own right, paralleling the explosion in research on 2D materials since 2004 [17, 18]. As compared with graphene, GO offers much more flexibility in tailoring its material properties via manipulation of the oxygen-containing functional groups (OFGs) [19]. Most importantly, GO offers facile synthesis processes as well as a strong compatibility with large-scale manufacturable on-chip integration, enabled by chemical oxidation of graphite and subsequent exfoliation and self-assembly in solution [20, 21]. The marriage between integrated photonics and GO has led to the birth of integrated GO photonics, which has become a very active and fast-growing branch of on-chip integration of 2D materials in order to achieve novel functionality of integrated photonic devices.

Here, we review the advances in this interdisciplinary field, focusing on the remarkable optical properties of thin GO films. While GO has been the subject of previous reviews [18, 19, 22], these have mainly focused on its basic chemical, electronic, and optical properties and their related applications. Here, we focus on the opportunities arising from the innovative integration of GO films with photonic devices, highlighting methods for on-chip integration and their diverse optical applications.

The review is structured as follows. An introduction of the tunable bandgap and optical properties of GO is presented in Section 2. In Section 3, the methods for on-chip integration of GO films are reviewed and discussed. In Section 4, we summarize recent work on functional integrated photonic devices incorporated with GO, being categorized into either passive (linear and nonlinear) or active (electrically interfaced) devices. The current challenges and future perspectives of integrated GO photonics are elaborated in Section 5, followed by conclusions in Section 6.

## 2. Optical properties of GO

As one of the most important derivatives of graphene, GO contains different OFGs on the basal plane and sheet edges, such as epoxy, hydroxyl, carbonyl and carboxyl groups, as illustrated in **Fig. 1(a)** [19, 23-26]. The type and degree of coverage of the OFGs in GO are variable, primarily depending on the different preparation processes [19]. **Fig. 1(b)** shows transmission electron microscopic (TEM) images of a single suspended GO sheet. Different graphitic areas are indicated by different colors in the right image. As can be seen, GO exhibits a highly inhomogeneous structure [27]. The disordered structure of GO can also be observed in the scanning tunneling microscopic (STM) image shown in **Fig. 1(c)**. Compared with the crystalline order of highly oriented pyrolytic graphite (HOPG) (inset on the left bottom), the STM image of GO shows a rough surface, featuring a peak-to-peak topography of ~1 nm caused by the OFGs and defects [28].

Owing to its heterogeneous atomic structure, GO exhibits a series of distinctive material properties. Compared with graphene, which consists entirely of $sp^2$-hybridized carbon atoms, GO is a 2D network consisting of variable $sp^2$ and $sp^3$ concentrations, thus enabling interesting and tunable electrical, chemical, and optical properties. For example, GO has excellent field emission properties, which are promising for field-effect transistors (FETs) [29, 30]. The high flexibility in tuning its conductivity and pseudo-capacitance also makes GO a good candidate for high performance transparent conductive electrodes and supercapacitors [31-33]. In addition, thanks to the large surface area and ample OFGs, GO can bind various catalysts or active materials for hydrogen storage and generation [34]. The excellent electrical and chemical properties of GO have been reviewed previously [17-19]. Here, we mainly focus on the prominent optical properties of GO as well as their broad photonic applications.

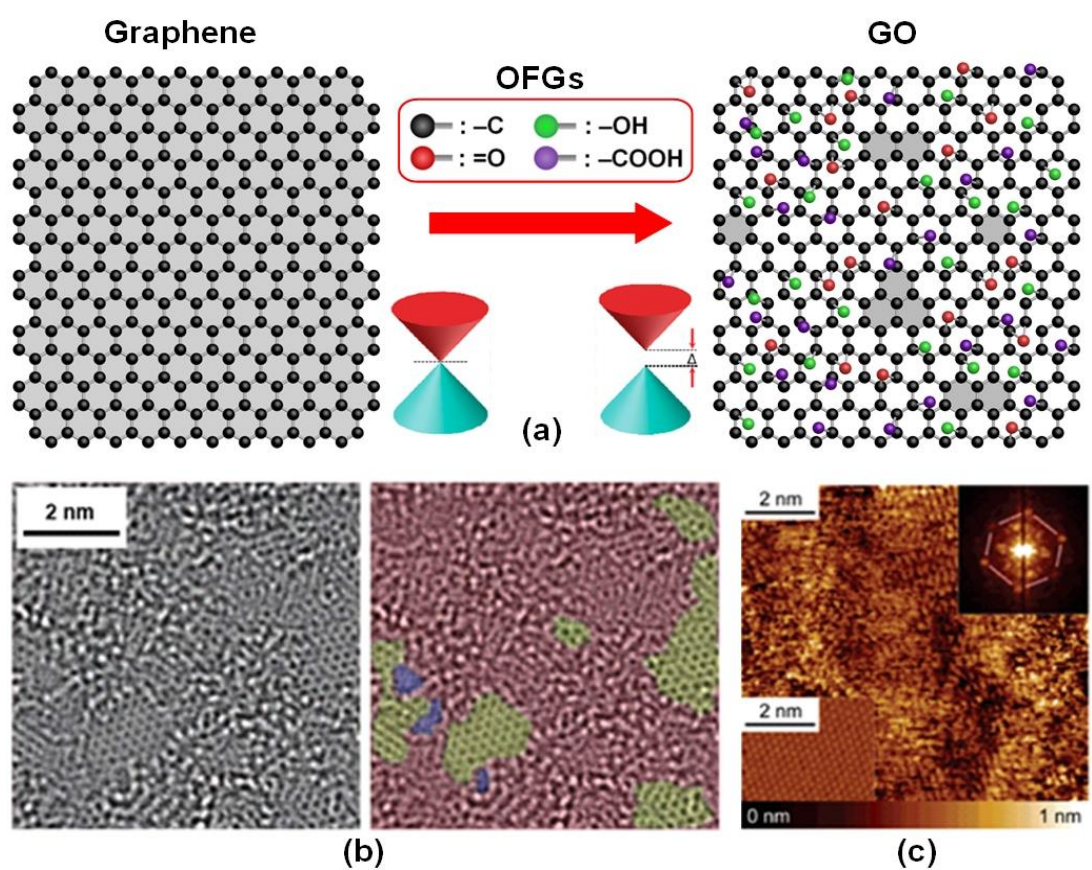

Figure 1. Atomic structure of GO. (a) Schematics of atomic structures and bandgaps of graphene and GO. (b) TEM images of a single suspended sheet of GO. On the right, holes are indicated in blue, graphitic areas in yellow, and OFG regions in red [27]. (c) STM image of a monolayer GO on a highly oriented pyrolytic graphite (HOPG) substrate. Insets on the right top and left bottom show the Fourier transform of the image and the STM image of HOPG under identical conditions, respectively [28].

## 2.1 Large and tunable optical bandgap

In contrast to graphene that has a metallic behavior with zero bandgap, GO features both conducting π-states from $sp^2$ carbon sites and a large energy gap between the σ-states of its $sp^3$-bonded carbons [19]. Pristine GO is a dielectric with a typical bandgap > 2 eV that can be readily tailored by tuning the ratio of the $sp^2$ and $sp^3$ fractions via reduction or doping treatments [20, 35]. This forms the basis for manipulating GO's material properties such as the conductivity, refractive index, and absorption, which enables a wide range of applications [19].

Based on the local-density approximation calculation, Yan *et al.* [36] verified that the bandgap of GO can be tuned from 4 eV to a few tenths of an eV by changing the coverage of the OFGs (**Fig. 2(a)**). In Ref. [21], Yang *et al.* used femtosecond pulsed laser reduction to tune the bandgap of GO films, achieving a tunability from ~2.4 eV (as-prepared GO) to ~0.1 eV (complete reduction) when increasing the laser power (**Fig. 2(b)**). Similar results have also been reported by Guo *et al.*[37], where they further fabricated a bottom-gate GO FET and obtained an optimized on−off ratio of 56 by tuning the laser power.

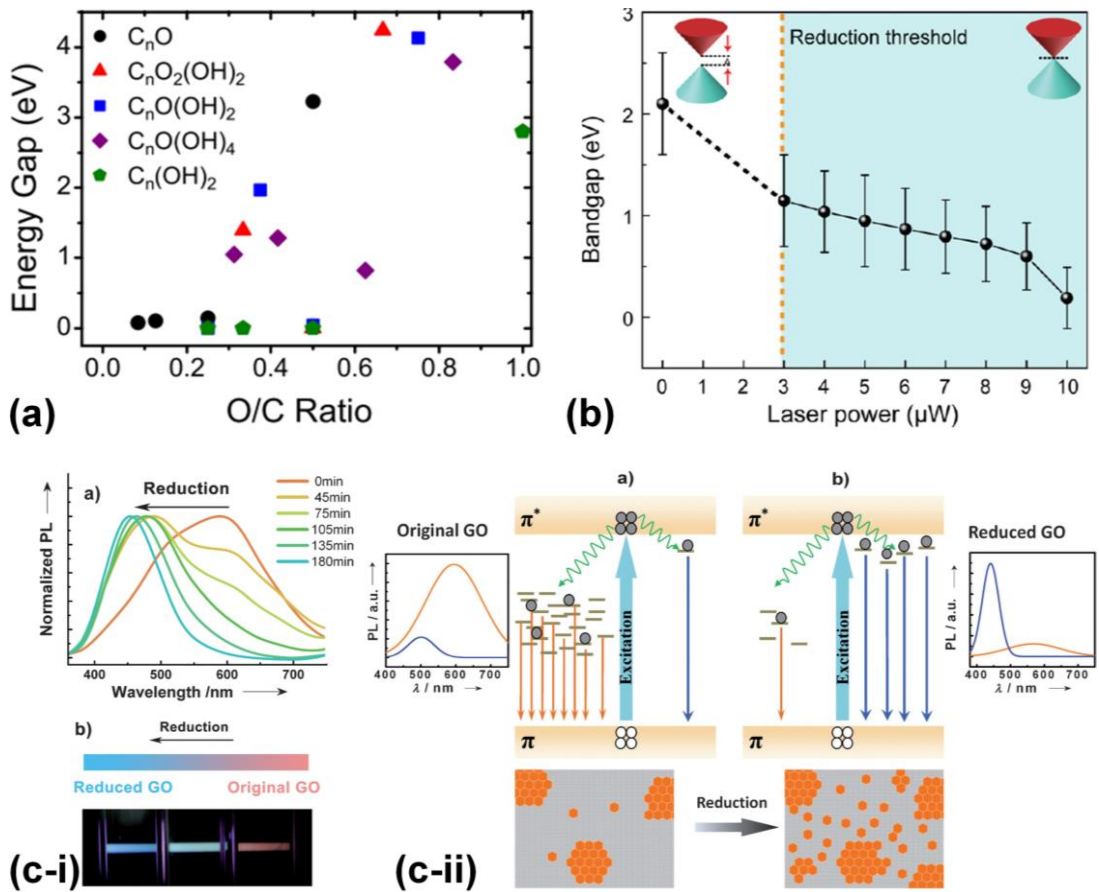

Figure 2. Tunable bandgap of GO. (a) Energy gap as a function of the overall oxygen-to-carbon (O/C) ratio for GO with different amounts of $sp^2$ carbon (C), epoxide ($C_2O$), and the 1,2-hydroxyl pair ($C_2(OH)_2$) [36]. (b) Tunable bandgap of GO as a function of laser power [21]. (c) Tunable PL from GO [38]: (i) PL spectra after different exposure times (0–180 min) to photothermal reduction treatment and photographs of PL emission at 0 min (yellow-red), 75 min (green), and 180 min (blue); (ii) schematic of the PL emission mechanism.

The large and tunable direct bandgap of GO results in efficient broadband photoluminescence (PL) in the near-infrared, visible and ultraviolet wavelength regions [19, 26, 39]. This contrasts with both silicon that has an indirect bandgap and graphene, where its zero bandgap yields no PL unless assisted by phonons. There are two main PL bands for GO, one in the blue region near 430 nm and the other in the longer visible wavelength range of

500 nm – 650 nm. By changing the bandgap of GO, tunable PL (**Fig. 2(c)**) has been demonstrated [38-40].

The reduction of GO has been widely used for changing its bandgap [19, 20]. Pristine GO sheets contain mostly $sp^3$ domains and fewer $sp^2$ domains, while the fraction of $sp^2$ domains increases with the degree of reduction. Fully reduced GO has material properties that are very close to graphene [21], providing a new way of fabricating high quality graphene-like films cost-effectively and on a large scale. The reduction of GO films can be achieved using thermal, chemical, or photo reduction methods [18, 22, 26]. Although the first two have a strong capability for removing OFGs, they are not particularly compatible with on-chip integration and so are not the focus of this article. In contrast, photo reduction does not involve high temperatures or toxic chemicals and offers unique advantages due to its moderate reaction conditions, exquisite control over the degree of reduction, and capability of advanced patterning designs. It can also be used to post-process films in-situ after they are integrated onto chips. Photo reduction can be further classified into three categories: photothermal, photochemical, and laser-induced localized reduction [20]. Here, we treat laser reduction as a distinct approach due to its unique ability for flexible and in-situ patterning, although it also involves some degree of photochemical and/or photothermal processes.

Ultrafast direct laser writing (DLW) has been widely used as a laser reduction method that is capable of both 2D and 3D writing of arbitrary patterns [41-48]. Ultrashort laser pulses offer low thermal effects that enable high fabrication resolution as well as rich light-matter interaction mechanisms and dynamics. The laser writing process is non-contact and mask-free, yielding flexible and fast prototyping with reduced fabrication cost and improved efficiency. **Fig. 3(a)** shows the fabrication of thin GO flat lenses via DLW [42, 49]. The concentric rings were fabricated by DLW to convert GO to reduced GO (rGO) via photo reduction. The controllable removal of the OFGs leads to a continuously controllable reduction in film thickness and increase in refractive index, accompanied by a small decrease in transmission.

Another important laser reduction method that is particularly suitable for the fabrication of periodic patterns is based on laser interference. This method can pattern large areas in a short period of time [50], and by selecting the appropriate processing parameters it can create diverse surface topographies. **Fig. 3(b)** shows GO gratings with a period of ~2 µm fabricated with a two-beam laser interference system [64]. By changing the angle between the two laser beams, the periodicity of the gratings could be tuned within a certain range.

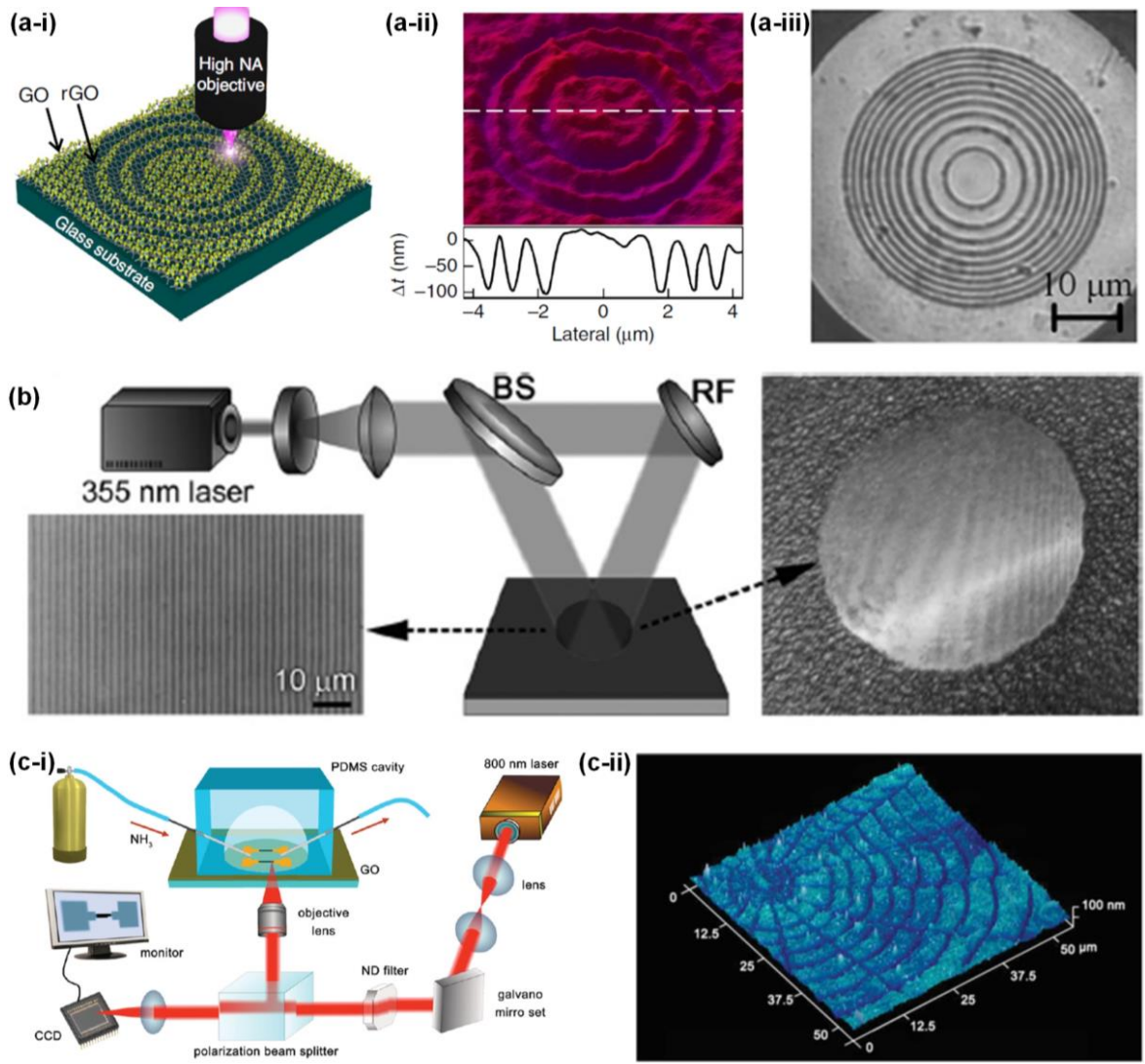

Figure 3. Laser processing of GO films. (a) Fabrication of GO flat lenses via DLW: (i) schematic illusion of DLW [42]; (ii) – (iii) topographic profile [42] and microscopic image [49] of the fabricated GO lenses, respectively. (b) Schematic illustration of a two-beam laser interference system and images of the fabricated GO gratings [51]. (c) Laser-assisted doping of GO [35]: (i) system schematic; (ii) 3D atomic force microscopic (AFM) images of N-doped rGO micropattern.

For both DLW and the laser interference, the accompanying laser heating caused by irradiation can sometimes trigger local chemical/physical reactions in GO, such as the breaking of the oxygen-containing bonds, creating defects by taking some of the carbon atoms, and forming new chemical bonds with molecular compounds in the local ambient environment. The transformation from GO to rGO via laser irradiation generally follows two basic processes: i) photochemical removal of oxygen from the GO surface, sometimes accompanied by laser ablation; ii) structural reorganization of the newly formed, reduced, carbon lattice into the planar, hexagonal, $sp^2$-conjugated graphene structure. As a result, changes in the film material properties after these processes are permanent and irreversible. During the reduction of GO under laser irradiation, simultaneous doping of the rGO can be realized in a dopant precursor environment by taking advantage of defect sites in GO. By using DLW in a controlled ammonia environment, Guo *et al.* [35] demonstrated efficient reduction of GO to N-doped graphene, as shown in **Fig. 3(c)**.

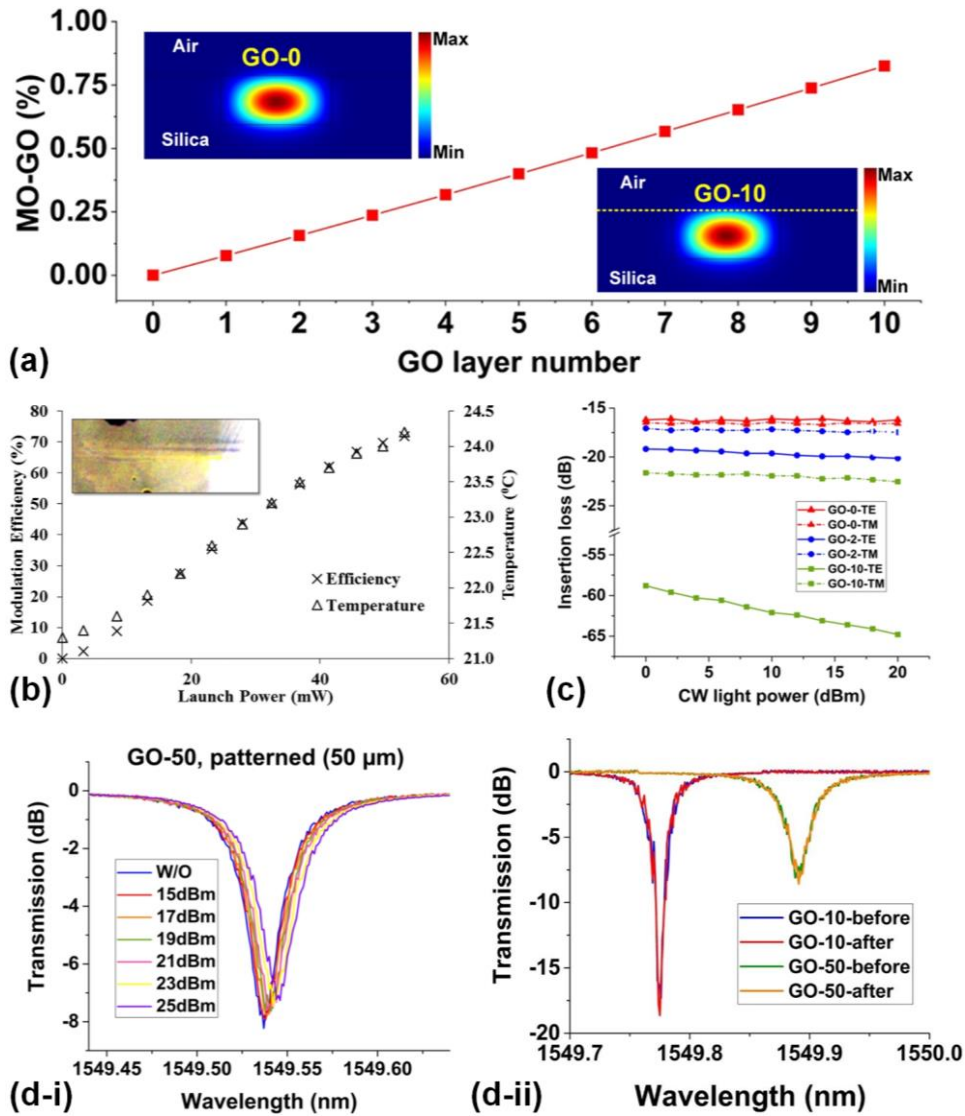

Figure 4. Photo-thermal reduction of GO in GO-coated optical waveguides. (a) Mode overlap with GO versus layer number for GO-SiN hybrid waveguides (cross section: 1.6 μm × 0.66 μm). Insets show TE mode profiles for the SiN waveguides without GO and with 10 layers of GO. (b) Modulation efficiency versus control laser power and corresponding temperature change measured near a GO-coated polymer waveguide. Inset shows the darkening of the GO film induced by photo-thermal reduction [52]. (c) Measured TE and TM polarized insertion losses versus input CW power for GO-coated doped silica waveguides [53]. (d) Photo-thermal reduction of GO in GO-coated microring resonators (MRRs) [54]: (i) power-dependent transmission spectra of a doped silica MRR with a patterned GO film measured using a pump-probe method; (ii) transmission spectra before tuning on and after tuning off high-power pump.

For integrated waveguides coated with GO films, the intensity of the evanescent field out of the waveguides that interacts with the films is much lower than what is present during DLW and laser interference (**Fig. 4(a)**), thus leading to a relatively weak light-matter interaction. In this case, reversible changes in the GO material properties via photo-thermal reduction can be observed [52-54], where the laser power generates heat and increases the temperature of the hybrid waveguides (**Fig. 4(b)**), temporarily modifying the OFGs. Photo-thermal induced changes in the OFGs can modify the material properties such as linear loss (**Figs. 4(c)**) and Kerr nonlinearity ($n_2$), depending on factors such as the average light power and mode overlap. The time response for photo-thermal changes is relatively slow (typically on millisecond timescales [52]), which is distinct from two photon absorption (TPA)-induced loss that occurs near-instantaneously and depends on the peak power. Thermally reduced-GO films are typically unstable and can easily oxidize in oxygen-containing environments after tuning off the laser power (**Figs. 4(d)**).

*2.2 Linear optical properties*

Due to the existence of OFGs, GO exhibits distinct linear optical properties as compared to graphene. **Fig. 5(a)** shows the measured linear refractive index *n* and extinction coefficient *k* of a GO film in the wavelength range of 200 nm – 25 μm [55]. **Fig. 5(b)** compares the *n*, *k* of GO and graphene films measured by spectral ellipsometry [56]. As can be seen, GO films exhibit a high refractive index of about 2 in the wavelength range covering visible, near-infrared, and mid-infrared regions. On the other hand, owing to its comparatively large material bandgap, GO has an extinction coefficient that is much lower than graphene, particularly at near-infrared wavelengths – in fact, about two orders of magnitude lower. **Fig. 5(c)** shows the changes in *n*, *k* of GO with laser reduction power [21]. Both show a trend towards those of graphene as the power increases. The large dynamic tuning ranges in *n* and *k* also form the basis for efficient phase and amplitude modulation in photonic devices.

Similar to other 2D materials such as graphene and TMDCs [11, 15, 57], 2D layered GO films have a huge anisotropy in optical absorption, with significantly higher in-plane rather than out-of-plane absorption and the difference between them decreasing with layer number (**Fig. 5(d)**). This reflects the transition of the film properties toward a bulk (isotropic) material for thick films. Moreover, the bandwidth of the optical anisotropy of GO films is very broad − extending from visible to infrared wavelengths. These optical properties can be used for implementing broadband polarization selective devices with high polarization selectivity [53, 55, 58]. It is also interesting to note that even for thick films (e.g., >100 nm) beyond typical thicknesses for 2D materials, the intrinsic film loss anisotropy is still large enough to enable some polarization-dependent devices [58, 59].

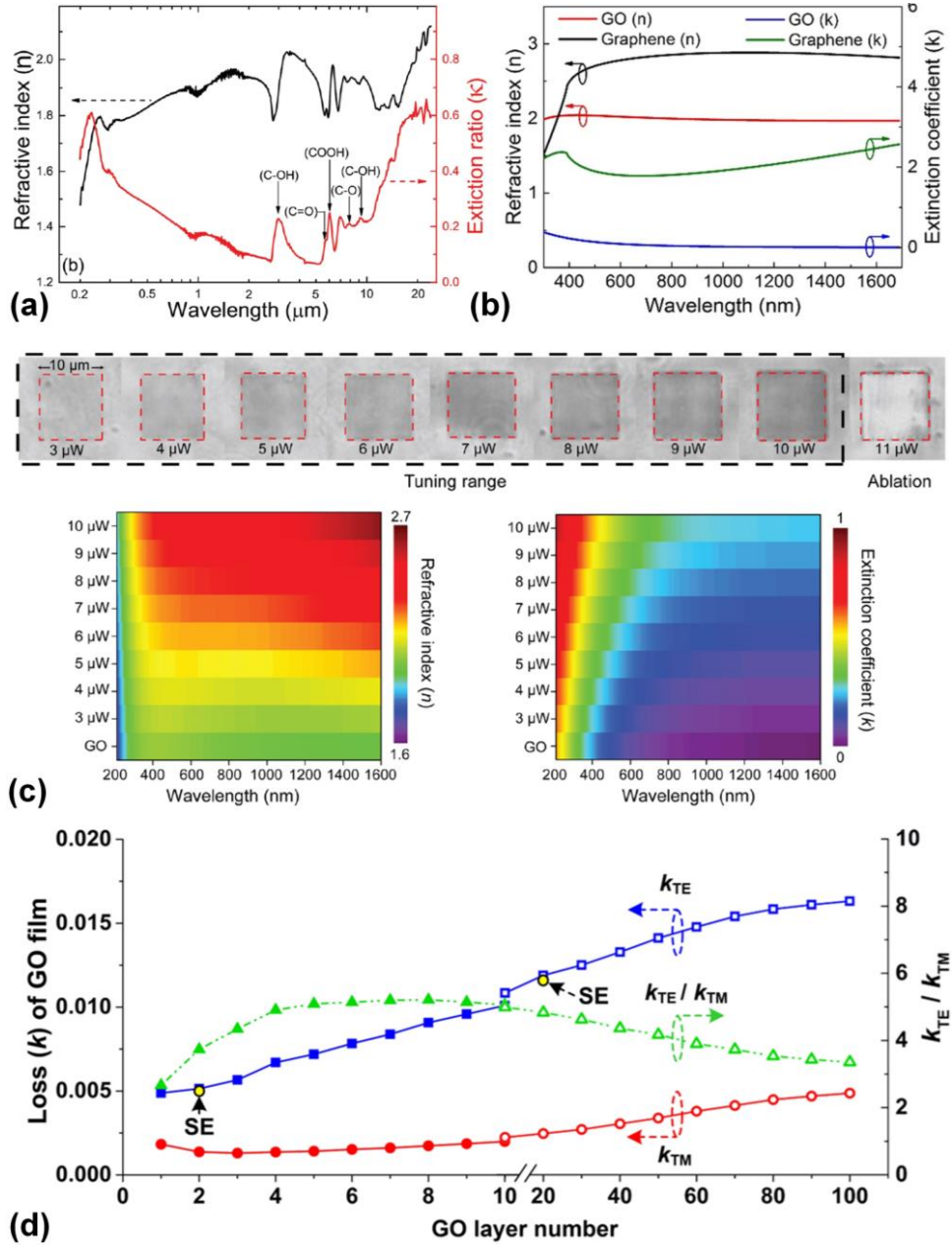

Figure 5. Linear optical properties of GO. (a) Measured refractive indices ($n$) and extinction coefficients ($k$) of GO from the visible (down to 200 nm) to mid-infrared (up to 25 μm) wavelengths [55]. (a) Comparison of $n$, $k$ between GO and graphene [56]. (c) Micrographs of various GO reduction levels (achieved with different laser powers) and the corresponding changes in $n$, $k$ [21]. (d) Material loss of the GO films for TE ($k_{TE}$) and TM ($k_{TM}$) polarizations as well as their ratio ($k_{TE} / k_{TM}$) [53].

## 2.3 Nonlinear optical properties

Aside from its interesting linear optical properties, the 2D nature of GO films combined with their tunable optical bandgap result in prominent nonlinear optical properties [60], featuring strong nonlinear optical absorption as well as a large Kerr nonlinearity.

The nonlinear optical absorption of GO, or the change in absorption with light intensity, can manifest in a number of different forms, from saturable absorption (SA, with absorption decreasing with light intensity) or reverse saturable absorption (RSA, with absorption increasing with light intensity), depending on the excitation wavelength and bandgap of the particular GO film. SA is widely used for pulse compression, Q-switching and mode locking [61, 62], whereas RSA is useful for optical limiting and high-power laser damage prevention [63, 64]. The SA of GO is mainly induced by ground-state bleaching of the $sp^2$ related states [65] that have a narrow energy gap of ~0.5 eV, where the optical absorption by electrons can easily be saturated, depleting the valence band and filling the conduction band [18, 66]. On the other hand, the RSA of GO is mainly caused by excited-state absorption (ESA) and TPA arising from the extended π-conjugate system that is typical for carbon-based materials [18, 26]. Since the sources for SA and RSA are different, they could possibly coexist in practical materials, thus leading to highly complex and wavelength dependent nonlinear absorption.

The Kerr optical nonlinearity, describing four-wave mixing (FWM), self-phase modulation (SPM), cross-phase modulation (XPM) and other effects [12, 14], has formed the basis of all-optical signal generation and processing with superior performance in speed and bandwidth than electronic approaches [67-69]. **Table 1** summarizes reported values of Kerr coefficient $n_2$ for GO films and other 2D materials. Here we focus on thin solid films. We also note that the Kerr nonlinearity of GO and GO nanocomposite, with the samples being dispersed in solutions, has been studied in Refs. [70-72]. The ultrahigh $n_2$ of GO films − about 4 to 5 orders of magnitude higher than that of silicon [14] − highlights their strong Kerr nonlinearity for many nonlinear optical applications such as FWM, SPM, XPM, third harmonic generation, and stimulated Raman scattering [67-69, 73].

The Z-scan method has been widely used for characterization of nonlinear absorption and nonlinear refractive index of GO films. Strong TPA in GO suspensions was observed by Liu *et al.* [74] via open-aperture Z-scan measurements at 532 nm using picosecond laser pulses, while ESA in GO was observed by using nanosecond laser pulses. In Ref. [75], Jiang *et al.* observed strong nonlinear optical limiting behavior in GO thin films via Z-scan measurements at 800 nm and 400 nm (**Fig. 3(a)**). They also found that the optical limiting performance was significantly enhanced upon partial reduction with laser irradiation or chemical reduction. Furthermore, GO was found to transit from SA to RSA when increasing the laser power at 800 nm.

**Table 1. Comparison of measured $n_2$ of GO films and other 2D materials. FWM: four-wave mixing. SPM: self-phase modulation. WG: waveguide. MRR: microring resonator.**

| Material | Wavelength (nm) | Film thickness | $n_2$ (×10$^{-14}$ m$^2$/W) | Method | Ref. |
|---|---|---|---|---|---|
| GO | ~800 | ~2 µm | ~70 | Z-scan | [65] |
| EGO[a] | ~800 | ~300 nm | ~5.7-36.3 | Z-scan | [76] |
| GO | ~1550 | ~1 µm | ~4.5 | Z-scan | [77] |
| GO | ~1550 | ~4 nm | ~1.5 | FWM in WG | [56] |
| GO | ~1550 | ~2−100 nm | ~1.2−2.7 | FWM in MRR | [54] |
| GO | ~1550 | ~2−20 nm | ~1.28−1.41 | FWM in WG | [78] |
| GO | ~1550 | ~2−40 nm | ~1.22−1.42 | SPM in WG | [79] |
| Graphene | ~1550 | 5–7 layers | −8 | Z-scan | [80] |
| Graphene | ~1550 | 1 layer | -10 | SPM in WG | [81] |
| MoS$_2$ | ~1064 | ~25 µm | 0.0188 | Z-scan | [82] |
| WS$_2$ | ~1040 | 57.9 nm | 0.0366 | Z-scan | [83] |
| BP | ~800 | ~30−60 nm | 68 | Z-scan | [84] |
| BiOBr | ~1550 | 140 nm | 3.824 | Z-scan | [85] |
| PdSe$_2$ | ~800 | ~8 nm | −0.133 | Z-scan | [86] |

a) EGO: electro-chemically derived graphene oxide

The Kerr nonlinear response of GO films has been measured by the Z-scan method at both 800 nm and 1550 nm [65, 77], obtaining a very high $n_2$ of $10^{-13} - 10^{-14}$ m$^2$/W. A tunable Kerr nonlinearity has also been achieved by changing the degree of reduction. **Fig. 3(b)** shows $n_2$ and nonlinear absorption coefficient $\beta$ of thin GO films (2-µm thick) at 800 nm. By continuously increasing the laser power, four stages (I − IV) of different nonlinear optical behaviors were observed. In addition to a tunable nonlinear optical response, a reversal in the sign of $n_2$ and $\beta$ was also observed during the transition from GO to rGO. **Fig. 3(c)** presents $n_2$ and $\beta$ for a 1-µm-thick GO film at 1550 nm, showing that $n_2$ is smaller than that at 800 nm, although still 4 orders of magnitude larger than that of silicon.

As compared to pristine GO, functionalized GO hybrid materials have the potential to further enhance the optical nonlinearity. Recently, Fraser *et al*. [87] proposed a flexible method to functionalize GO films with gold nanoparticles (AuNPs), resulting in hybrid GO-

AuNP films that displayed significantly enhanced nonlinear absorption and refraction (**Figs. 3(d)**). This can be attributed to the efficient energy and/or charge (electron) transfer upon photoexcitation, together with synergistic coupling effects between GO and the AuNPs. Similarly, an enhanced optical nonlinearity has been achieved through covalent functionalization of GO with other materials such as $Fe_3O_4$ [72] and fullerenes [88]. A significant enhancement in $n_2$ has also been reported [76] in GO films synthesized with an electrochemical method, which simultaneously yielded significantly improved material stability under high-power laser illumination (up to 400 mJ/cm$^2$, **Fig. 3(e)**) – useful for high-power nonlinear optical applications.

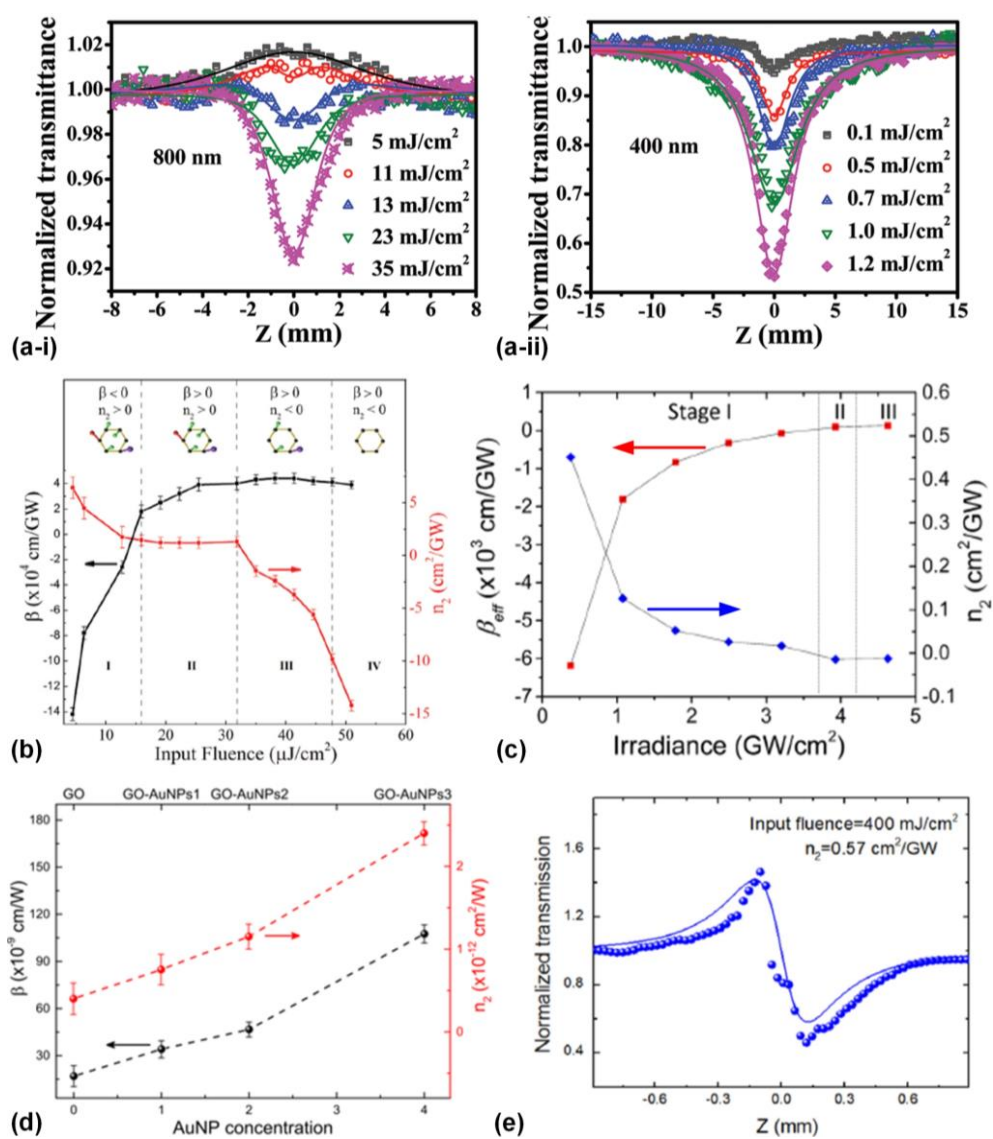

Figure 6. Nonlinear optical properties of GO films measured by the Z-scan method. (a) Open aperture Z-scan results of a GO film at (i) 800 nm and (ii) 400 nm under different laser fluences [75]. (b) Kerr coefficient $n_2$ and nonlinear absorption coefficient $\beta$ of GO at 800 nm as a function of laser fluence. Four different stages (I − IV) are labeled [65]. (c) $n_2$ and $\beta$ of GO at 1550 nm as a function of laser irradiance [77]. (d) Comparison of $n_2$ and $\beta$ of the AuNP-only, GO, and GO-AuNPs samples [87]. (e) Close aperture Z-scan result under a high laser fluence of 400 mJ/cm$^2$ for a GO film synthesized via an electrochemical method [76].

## 3. Integration of GO films onto photonic chips

The ability to integrate 2D layered GO films onto photonic chips could significantly facilitate manufacturable hybrid integrated devices for commercial applications outside of the laboratory – something that has thus far eluded most 2D materials. Their facile synthesis processes and strong amenability for large-scale on-chip integration give them significant advantages over other 2D materials. Moreover, GO also has high flexibility in engineering its material properties via different reduction methods, making it useful for diverse applications. In this section, we review methods for on-chip integration of GO films. We focus on solid films that can be integrated on chip, in contrast to suspensions being dispersed in solutions [70-72]. **Fig. 7(a)** shows a typical fabrication process flow for the on-chip integration of GO films, using the silicon-on-insulator (SOI) platform [12] as an example, although GO films can readily be introduced into other integrated platforms (e.g., SiN and dope silica [14]) with similar fabrication processes.

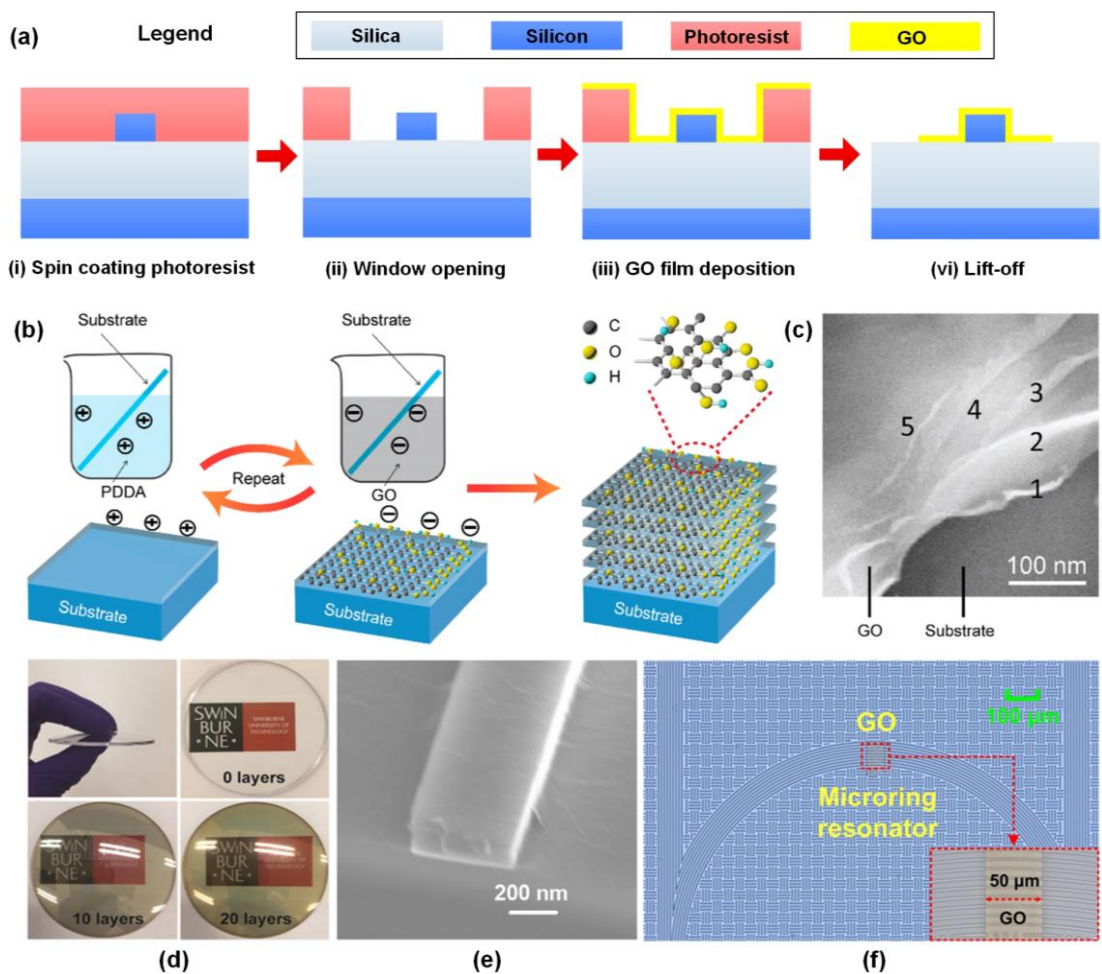

Figure 7. On-chip integration of GO films. (a) Typical fabrication process flow. (b) Schematic showing layer-by-layer GO film coating via self-assembly [21]. (c) SEM image of a 2D layered GO film (with up to 5 layers) [21]. (d) Large-area GO film coating on a silica substrate [21]. (e) GO film conformally coated on a silicon nanowire waveguide [79]. (f) Precise GO film patterning on an integrated MRR [54].

The deposition of thick GO films has typically been achieved by spray coating or drop-casting or methods [58, 89, 90], where the minimum film thickness is usually hundreds of nanometers. These approaches are mainly used for coating thick films in large areas, and so their applications to optical waveguides are limited to relatively large dimensions (> 5 µm, e.g., polymer waveguides) where there are no stringent requirements for the film thickness, uniformity, placement, and size.

The deposition of ultrathin 2D layered GO films can be achieved via a solution-based method that yields large-area, transfer-free, and layer-by-layer deposition [21, 53]. First, a high-quality GO solution is prepared by the chemical oxidation of graphite using a modified Hummers method [91] and vigorous sonication, where monolayer GO nanoflakes in a highly intact carbon framework with minimal residual impurity concentrations are obtained. This is quicker, safer, and more efficient than other GO synthesis methods such as the Brodie method that requires potassium chlorate and fuming nitric acid [26]. After preparing the GO solutions, four steps for the in-situ assembly of monolayer GO films are repeated to construct multilayer films on a target substrate, as shown in **Fig. 7(b)**. This GO film coating method can achieve precise control of the film thickness with an ultrahigh resolution of ~2 nm (i.e., the thickness for 1 layer). In contrast to the imprecise, largely unrepeatable, and unstable approach of mechanical layer transfer processes (e.g., using scotch tape) that have been widely used for other 2D materials such as graphene and TMDCs [5, 15, 92], this approach can be scaled up for manufacturing with highly controllable, repeatable, and stable fabrication processes. Unlike normal GO materials that are soluble in the water environment, the self-assembled films have high water resistance due to the solution-based synthesis and the adhesion between adjacent layers by electrostatic forces. The coated films can easily be removed by plasma oxidation, making it possible to reuse the integrated devices.

**Figs. 7(c)** and **(d)** show the images for the self-assembled GO films coated on silica substrates with 2D layered structure and high uniformity in large areas. For nanowire waveguides in **Fig. 7(e)**, this coating method can yield conformal coating allowing direct contact and enclosing of GO films with integrated photonic devices, which has rarely been achieved for other 2D materials. The conformal coating is very useful for efficient light-matter interaction but has been challenging to achieve for mechanical transfer approaches.

To accurately control the placement and size of the films for integrated photonic devices, patterning can be achieved with standard lithography and lift-off processes [93, 94]. The chip is first spin-coated with photoresist and then patterned via either photo or electron-beam lithography to open windows in the photoresist, using alignment markers to accurately control the window position [95, 96]. GO films are then deposited on the chip using the transfer-free, layer-by-layer coating method discussed above, and then patterned via lift-off processes. **Fig. 7(f)** shows a microscopic image of an integrated MRR with a 50-µm-long patterned GO film. By using e-beam lithography, a short pattern length of ~150 nm was achieved for 2 layers of GO [53], highlighting the achievable high patterning resolution. The layer-by-layer coating along with lithography and lift-off processes allows the precise control of the film thickness, placement, and size on integrated devices. Combined with the large-area, transfer-free coating nature, this method enables cost-effective, large-scale, and highly precise integration of 2D layered GO films on chip, representing a significant advance towards manufacturable integrated photonic devices incorporating 2D materials.

## 4. Integrated photonic devices incorporated with GO films

The superior optical properties and strong capability for large-scale, highly precise on-chip integration of GO films have enabled functional integrated photonic devices for a variety of applications in optical sciences. In this section, we review integrated photonic devices incorporated with GO films, being structured as follows. In **Sections 4.1** to **4.5**, we review passive photonic devices, including both linear (**4.1** to **4.4**) and nonlinear (**4.5**) devices. Subsequently in **Sections 4.6** and **4.7**, we review active (electrically interfaced) photonic devices, including light emitting devices (**4.6**) and photodetectors (**4.7**).

*4.1 Light absorbers*

High efficiency light absorbers play a critical role in photovoltaics [97], solar-thermal harvesting [98, 99], desalination [100], photodetection [101, 102], and concealment [103]. To implement light absorbers, different schemes have been used based on plasmonic metamaterials [104], dielectric gratings [105], hyperbolic metamaterial nanoparticles [106], and layered structures composed of 2D materials [44, 48, 107]. Amongst them, ultrathin large-area 2D materials offer unique advantages to achieve broadband absorption of unpolarized light over a wide range of angles. Owing to their ease of preparation and flexibility in tuning their optical properties, GO films have been used to implement high performance light absorbers [44, 48, 100, 108].

Recently, Lin *et al.* [48] demonstrated a 90-nm-thick GO film absorber (**Fig. 8(a-i)**) with a high absorptivity of ~85% for unpolarized visible and near infrared light covering almost the entire solar spectrum (~300 nm – 2500 nm, **Figs. 8(a-ii)** and **(a-iii)**). Gratings formed by alternating GO and rGO coupled the light into waveguide modes to achieve light absorption over incident angles up to 60°, showing heating up to 160 °C in natural sunlight. The GO film deposition was achieved by using the transfer-free, layer-by-layer coating method and the reduction of GO was achieved by DLW. The strong light absorption in thin films with a large surface area and the very broad spectral / angular responses are highly desirable for solar thermal applications.

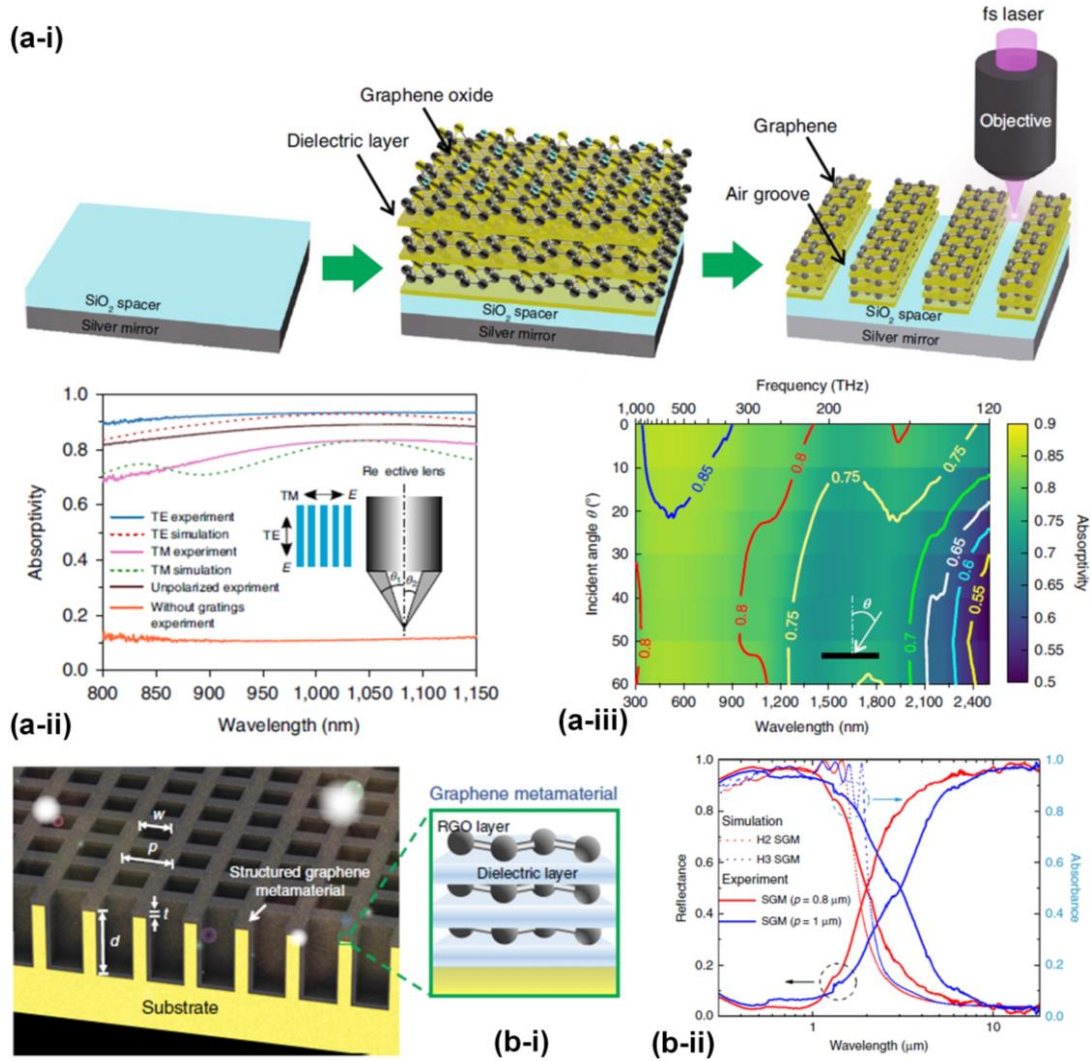

Figure 8. Light absorbers based on thin GO films. (a) 90-nm-thick GO gratings for strong and broadband absorption of unpolarized light [48]: (i) schematic of fabrication; (ii) – (iii) measured absorptivity spectra for different incident polarizations and angles, respectively. (b) A selective solar-thermal absorber based on a GO film coated on a metallic trench-like structure [44]: (i) device schematic; (ii) measured and simulated reflectance and absorbance spectra.

A solar-thermal absorber based on 3D structured rGO was reported subsequently [44], where the GO film (30-nm-thick) was conformally coated onto 3D metallic trench-like structure to form 3D resonant cavities (**Fig. 8(b-i)**), followed by photo-reduction to convert the GO into rGO. The wavelength selectivity of the resonant structure as well as the dispersionless and highly thermally conductive nature of rGO, resulted in an absorber with superior solar-selective and omnidirectional absorption. Further, it achieved a high solar-to-thermal conversion efficiency of 90.1% and a high solar-to-vapor efficiency of 96.2% (**Fig. 8(b-ii)**).

In 2014, Jiao *et al.* [109] demonstrated an approach to improving the efficiency of graphene/silicon Schottky-barrier solar cells by inserting a thin GO interfacial layer, achieving a dramatic improvement in the power conversion efficiency (PCE) of more than 100%. High efficiency rGO-silicon Schottky junction solar cells have also been reported [110], where the GO films were produced with a scalable vacuum filtration method and then reduced via thermal annealing. Chemical doping for different annealing temperatures and film thicknesses was found to increase the cell PCE by up to 220%. Recently, Nikolaos *et al.* [111] incorporated rGO nanoflakes in planar perovskite solar cells to obtain a high PCE of 13.6% − improved by 20% compared to reference devices. The rGO further stabilized the solar cells, which retained 40% of their initial PCE after 50 days of storage in a mildly humid, dark environment.

*4.2 Optical lenses and imaging devices*

Optical lenses are indispensable components in optical science and technology [42], and ultrathin flat lenses have revolutionized this field [42, 46, 49], achieving huge miniaturization of conventional lens systems in lab-on-chip devices, for example. Recent breakthroughs in nanophotonics have enabled ultrathin flat lenses based on metamaterials [112], metasurfaces [113], and super-oscillations [114], although their typically narrow operational bandwidth, complex designs, and time-consuming multi-step manufacturing processes have limited their practicality, especially for large-scale production.

The ability to widely tune the refractive index and absorption of thin GO films by laser photo-reduction has enabled the realization of nanometric flat lenses with 3D subwavelength focusing and accurate control of the wave-front [21, 42, 49, 115]. The excellent and robust focusing properties of these ultrathin GO film devices, together with their simple and scalable fabrication, have enabled the highly precise and efficient manipulation of optical beams.

A thin GO flat lens that can simultaneously manipulate the phase and amplitude of an incident beam was demonstrated in 2015 [42]. **Fig. 9(a-i)** shows the wavefront manipulation by the GO lens, which enables far-field 3D subwavelength focusing. The flat lens was fabricated by patterning sub-micrometer rGO concentric rings on a 200-nm-thick GO film using DLW. Broadband light focusing from visible to near infrared wavelengths (i.e., ~400 nm − 1500 nm, **Fig. 9(a-ii)**) was achieved, with an averaged absolute focusing efficiency of >32% over the entire band.

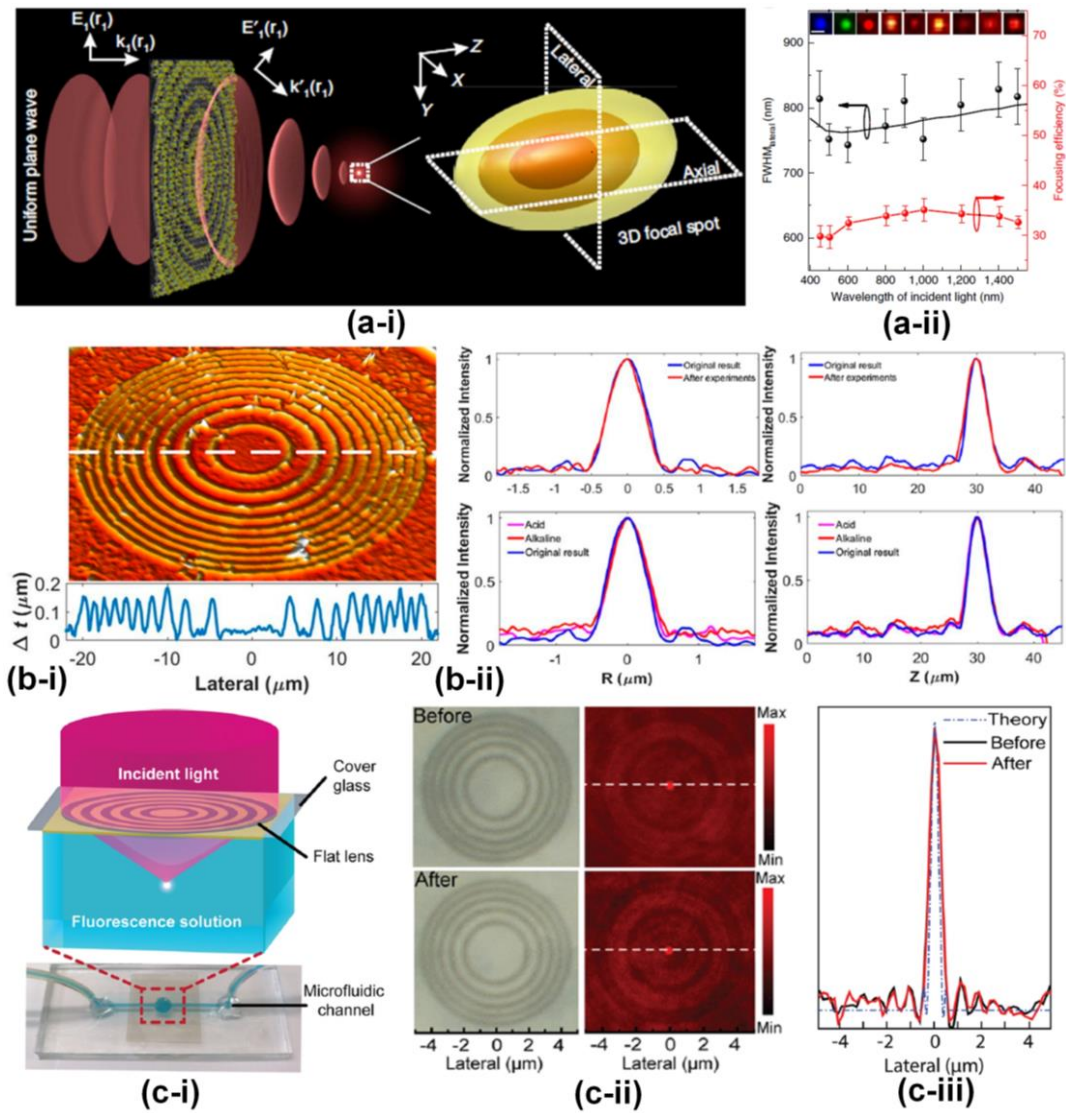

Figure 9. Flat lenses based on thin GO films. (a) A broadband GO flat lens with 3D subwavelength focusing [42]: (i) schematic of wavefront manipulation; (ii) broadband focusing ability from 450 nm to 1500 nm. (b) A rGO flat lens with high stability in harsh environments [49]: (i) topographic profile; (ii) performance comparison before and after ultraviolet exposure and strong acid/alkaline treatment. (c) A GO flat lens in a biocompatible microfluidic environment [21]: (i) image of device and schematic of operation principle; (ii) – (iii) microscope images and intensity distributions of the lens before and after one month immersion, respectively.

In 2018, the design of GO lenses based on the Rayleigh-Sommerfeld (RS) diffraction theory was proposed and experimentally verified [115]. In contrast to the classical Fresnel diffraction model under the paraxial approximation for low numerical-aperture (NA) focusing, the RS model allowed a more accurate design of GO lenses with arbitrary NA's and focal lengths. In 2019, Cao *et al.* [49] demonstrated a resilient rGO flat lens that maintained a high level of structural integrity and outstanding focusing performance in harsh environments (**Fig. 9(b)**), including low earth orbit space and strong corrosive chemical (pH = 0 and pH = 14) environments. Similarly, a thin GO flat lens with high robustness in biocompatible microfluidic environments (**Fig. 9(c)**) was also reported [21].

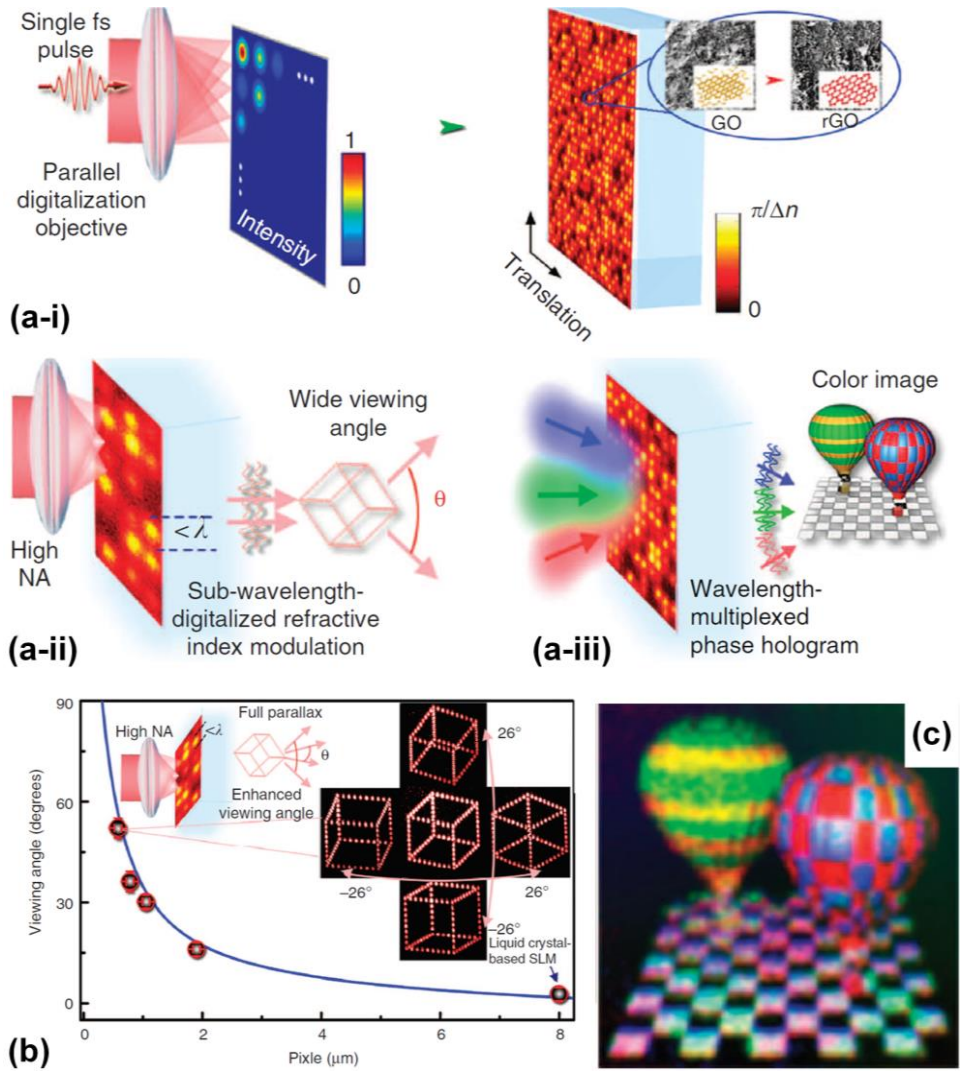

Figure 10. Photo reduction of GO for 3D holographic imaging [41]. (a) Schematic illustration of operation principle: (i) optical digitalization of refractive-index/phase modulation by reducing GO via femtosecond laser pulses; (ii) wide-angle 3D imaging achieved by confining the photoreduction at a subwavelength scale; (iii) reconstruction of color objects through wavelength-multiplexed phase hologram recorded in GO polymers (b) Viewing angle as a function of size of the constitutive pixel. (c) Reconstructed color objects.

In addition to optical lenses, GO also provides new ways of implementing versatile holographic components in display systems with its exceptional laser-tunable electronic and optical properties. 3D holographic displays were reported [41], where subwavelength-scale multilevel optical index modulation of rGO was achieved with femtosecond laser pulses, yielding wide-angle and full-color 3D rGO holographic images (**Fig. 10(a)**). When the pixel size was reduced to 0.55 µm, static 3D holographic images with a wide viewing angle of up to 52 degrees were achieved (**Fig. 10(b)**). In addition, the spectrally flat optical index modulation of the rGO films enabled wavelength-multiplexed holograms for full-color images (**Fig. 10(c)**). The large and polarization-insensitive phase modulation ($> \pi$) in rGO composites also allowed the restoration of vectorial wavefronts of polarization discernible images through vectorial diffraction of the reconstruction beam.

*4.3 Polarization selective devices*

Polarization selective devices are core components for polarization control in optical systems – a fundamental requirement for optical technologies [116-118]. Recently, the huge optical anisotropy and dispersionless nature of 2D materials such as graphene, GO, and TMDCs have been widely recognized and exploited to implement polarization-selective devices [11, 15, 119-122]. As compared with conventional polarizers based on polarization dependent mode overlap with lossy bulk materials [123, 124], the material absorption anisotropy of 2D materials provides a new way to further improve polarization selectivity. Moreover, the broadband response of 2D materials yields very large bandwidths (typically several hundred nanometers from the visible to infrared wavelengths), which are extremely challenging to achieve with silicon photonic polarizers [118, 125, 126].

Owing to its broadband high material anisotropy and ease of fabrication, GO has distinctive advantages for implementing polarization selective devices. In Ref. [58], a broadband GO-polymer waveguide polarizer with a high polarization dependent loss (PDL) of ~40 dB was reported, where the GO films (~2-μm-thick) were introduced onto an SU8 polymer waveguide using the drop-casting method. **Fig. 11(a)** show the device schematic and the top view of the drop-casted GO with a diameter of ~1.3 mm. In Ref. [53], Wu *et al.* demonstrated GO waveguide polarizers and polarization-selective MRRs based on a CMOS-compatible integrated platform. Doped silica waveguides and MRRs with both uniformly coated and patterned GO films were fabricated. A high PDL of up to ~53.8 dB was achieved for waveguide polarizers (**Fig. 11(b-i)**), and for GO-coated MRR polarizers, a polarization extinction ratio (ER) of up to ~8.3-dB was observed between the TE and TM resonances (**Fig. 11(b-ii)**). These GO-based polarizers have simpler designs with higher fabrication tolerance as compared with silicon photonic polarizers, which usually require precise design and control of the dimensions [117, 118].

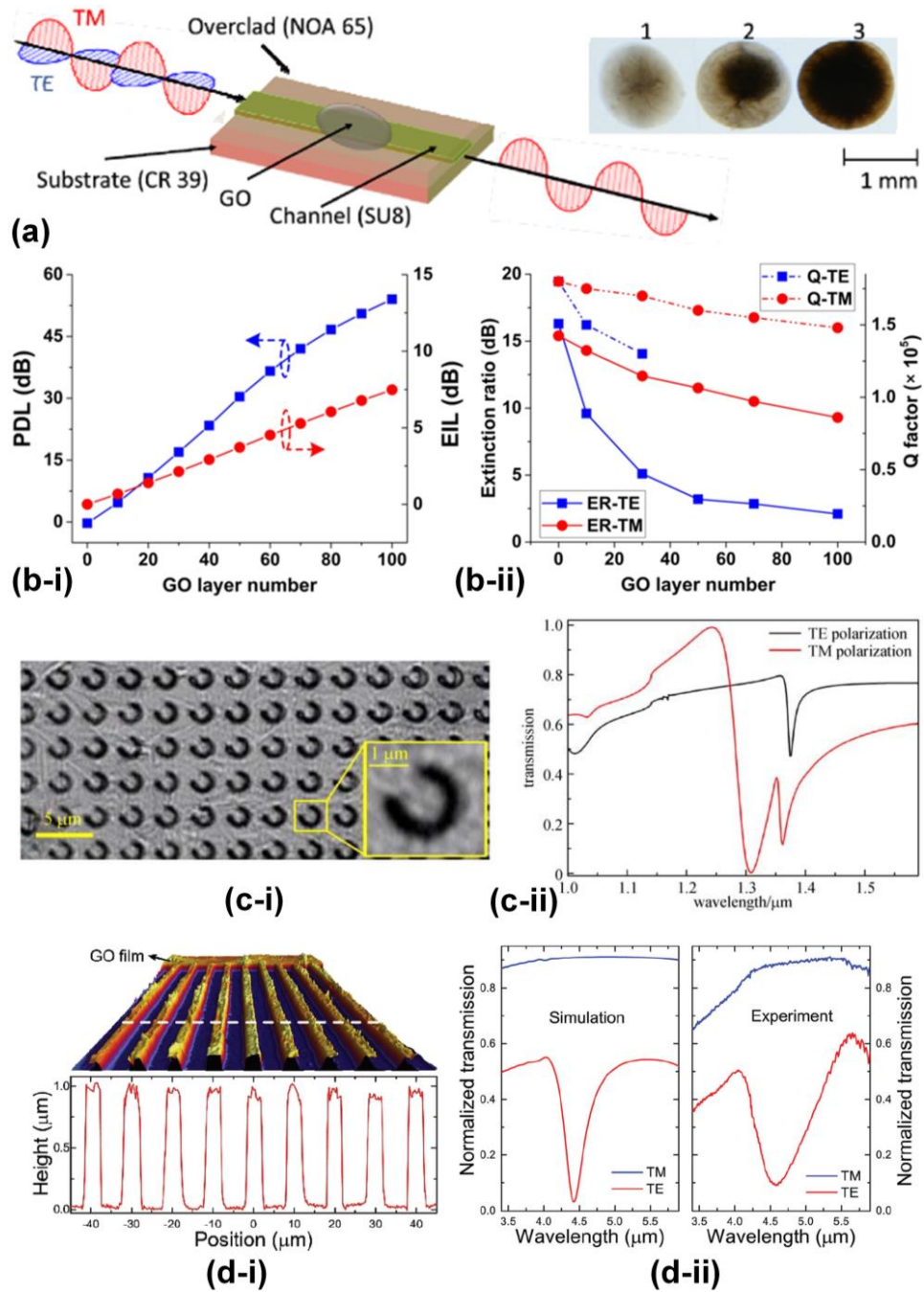

Figure 11. Polarization selective devices based on thin GO films. (a) Schematic of a GO-polymer waveguide polarizer and top view of drop-cast GO coating with different number of solution drops [58]. (b) Performance of GO-doped silica waveguide and MRR polarizers [53]: (i) polarization dependent loss (PDL) and excess insertion loss (EIL) for the waveguide polarizers; (ii) extinction ratios (ERs) and Q factors (Q) for the MRR polarizers. (c) GO thin film polarizer based on 2D periodic C-shape array [25]: (i) microscopic image of the fabricated device; (ii) transmission spectra with TE and TM polarized light incidence. (d) Mid-infrared thin film polarizer based on GO gratings [55]: (i) 3D topographic view of the fabricated device; (ii) theoretical and experimental transmission spectra for TE and TM polarizations.

In contrast to guided mode devices, thin film polarizers, where the optical beam is normal to the surface, are typically used for free space optical systems [127]. In 2017, a thin GO film polarizer based on periodic C-shape array (**Fig. 11(c-i)**) was demonstrated [25]. Due to the strong light confinement within the asymmetric C-shaped structure, the transmission spectra were highly sensitive to the incident polarization (**Fig. 11(c-ii)**). By optimizing the GO film thickness and the C-shape geometry, a high ER >3000 was achieved. Owing to the dispersionless nature of the GO film, the operation band can be tuned over a wide range from the visible (600 nm) to the near infrared (1.6 µm). Based on accurately characterizing the dispersion of GO from the visible (200 nm) to MIR (up to 25 µm) region, Zheng *et al.* [55] subsequently demonstrated high performing thin film polarizers on free-standing GO films (**Fig. 11(d-i)**), achieving a large ER (∼20 dB) and controllable working wavelengths in mid-infrared region (**Fig. 11(d-ii)**).

*4.4 Sensors*

GO films have been used as molecular sieves [128], enabled by their high surface-to-volume ratio, exceptional molecular permeation properties, and high adsorption capacity [129]. Highly sensitive detection of vapor phase volatile organic compounds (VOCs) has been demonstrated with a GO-coated silicon MRR (**Fig. 12 (a)**), with the detection sensitivity enhanced by a factor of 2 due to capillary condensation within the GO interlayers. GO films have also been used for tracking particles (**Fig. 12 (b)**). By characterizing the imaging relationship for a rGO flat lens, with an array of nanoholes that have micrometer spacing as the reference object, a high tracking accuracy of 10 nm was achieved.

Interestingly, despite being fluorescent, GO can also quench fluorescence. This quenching effect mainly results from fluorescence (or Förster) resonance energy transfer, or non-radiative dipole–dipole coupling, between the fluorescent species and GO [26, 130]. This has formed the basis of optical biosensors for sensing dye-labeled DNA and biomolecules. In 2010, He *et al.* [131] reported a GO-based multicolor fluorescent DNA nanoprobe that allowed rapid and selective detection of DNA targets, which is able to detect a range of analytes when combined with functional DNA structures. Liu *et al.* [132] subsequently designed a promising self-assembled homogeneous immunoassay for tracing biomarker proteins with distance independent quenching efficiency, based on modulating the interaction between GO sheets and inorganic luminescent quantum dots.

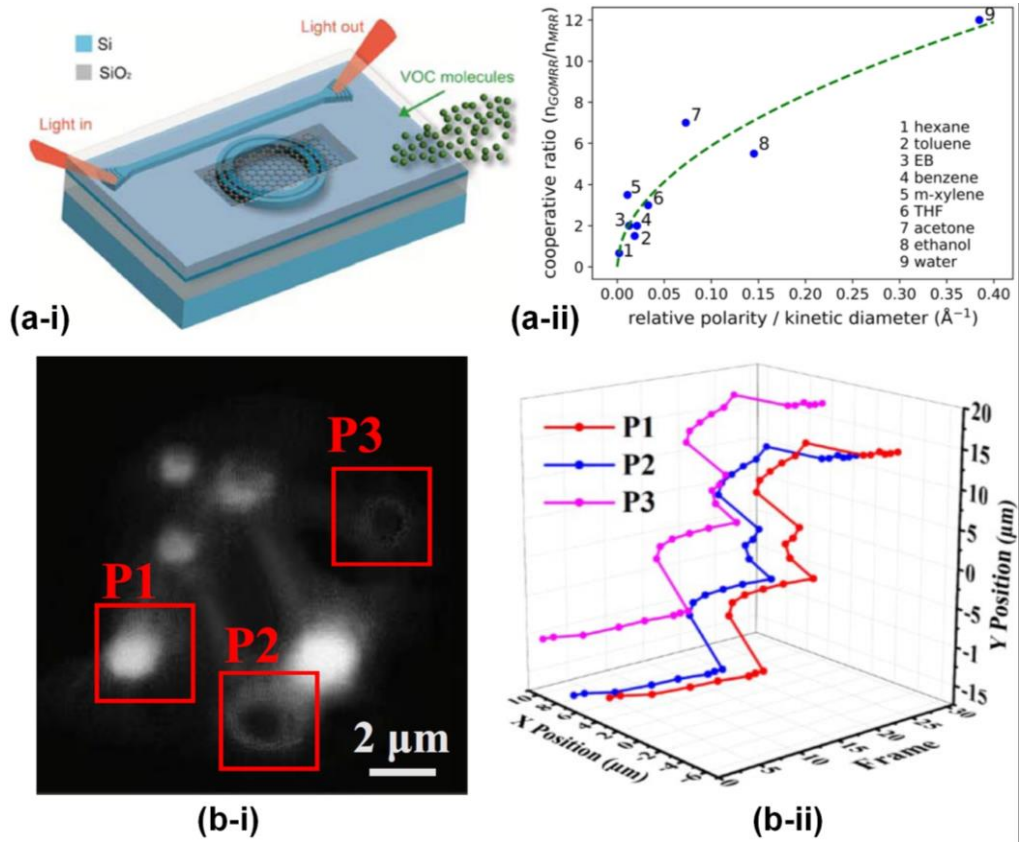

Figure 12. GO sensors. (a) Detection of volatile organic compound (VOC) molecules based on a GO-coated silicon MRR [129]: (i) device schematic; (ii) cooperativity ratio (nGOMRR/nMRR) as a function of the relative polarity – kinetic diameter ratio for all VOCs tested. (b) Particle nanotracking based on a rGO flat lens [133]: (i) image of the particles from the rGO lens; (ii) trajectories of the three particles labeled in (i) as a function of video frame number.

*4.5 Nonlinear optical devices*

Nonlinear integrated photonic devices based on the Kerr effect offer powerful solutions to generate and process signals all-optically, with superior processing speed compared to electronic devices, as well as the added benefits of a compact footprint, low power consumption, high stability, and the potential to significantly reduce cost by mass production [12, 14, 134]. Although silicon has been a leading platform for integrated photonic devices [12, 135, 136], its strong TPA at near-infrared telecommunications wavelengths poses a fundamental challenge for Kerr nonlinear devices operating in this wavelength region. Other CMOS compatible platforms such as SiN and doped silica [13, 14] have a much lower TPA, although they still suffer from intrinsic limitations arising from their comparatively low Kerr nonlinearity. The quest for high-performance nonlinear integrated photonic devices has become a driving force for integrating highly nonlinear materials onto chips to overcome limitations of the basic device platforms [7, 56, 137].

The giant Kerr nonlinear response of 2D layered materials such as graphene, GO, BP, and

TMDCs has been widely recognized and exploited for high performance nonlinear photonic devices offering new capabilities [7, 56, 60, 62, 65, 138-143]. GO has a number of unique advantages compared with other 2D materials, highlighted by a giant Kerr nonlinearity ($n_2$) that is about 4 orders of magnitude larger than silicon as well as a linear absorption that is over 2 orders of magnitude lower than graphene at infrared wavelengths [56]. An even more appealing feature is that the linear absorption of GO, unlike graphene, is not fundamental and can be reduced through optimizing film fabrication processes. Moreover, the large bandgap (> 2 eV) of GO yields low TPA in the telecommunications band [21, 37], which is highly desirable for Kerr nonlinear processes such as FWM and SPM. Finally, these advantages are on top of those already discussed, such as the capability for large-scale, highly precise integration, and flexibility in engineering the material properties by altering the OFGs through reduction methods.

FWM is a fundamental third-order nonlinear process that, in degenerate form, accounts for the Kerr effect (i.e., intensity dependent refractive index). It has been widely used for all-optical signal generation and processing, including wavelength conversion [144, 145], optical frequency comb generation [146, 147], optical sampling [137, 148], quantum entanglement [149, 150], and many others [73, 151, 152].

The use of 2D layered GO films to enhance the nonlinear optical performance of integrated photonic devices was first demonstrated by Yang *et al.* [56], where a net enhancement of up to ~6.9-dB in the FWM conversion efficiency (CE) was achieved for a ~1.5-cm-long doped silica waveguide uniformly coated with 2 layers of GO (**Fig. 13(a)**). The GO film, with a thickness of about 2 nm per layer, was introduced on top of the waveguide via the transfer-free, layer-by-layer coating method mentioned in **Section 3**. Enhanced FWM in GO-coated SiN waveguides was subsequently reported [78]. SiN waveguides with both uniformly coated (20-mm-long, 1 and 2 layers) and patterned (1.5-mm-long, 5 and 10 layers) GO films were fabricated (**Fig. 13(b-i)**). The maximum CE enhancement (~9.1 dB) was achieved for the waveguide patterned with 5 layers of GO (**Fig. 13(b-ii)**), reflecting the trade-off between CE enhancement and the loss increase in the hybrid waveguides. In addition to more significant enhancement of CE, patterning the films also yielded greatly broadened conversion bandwidth (**Fig. 13(b-iii)**).

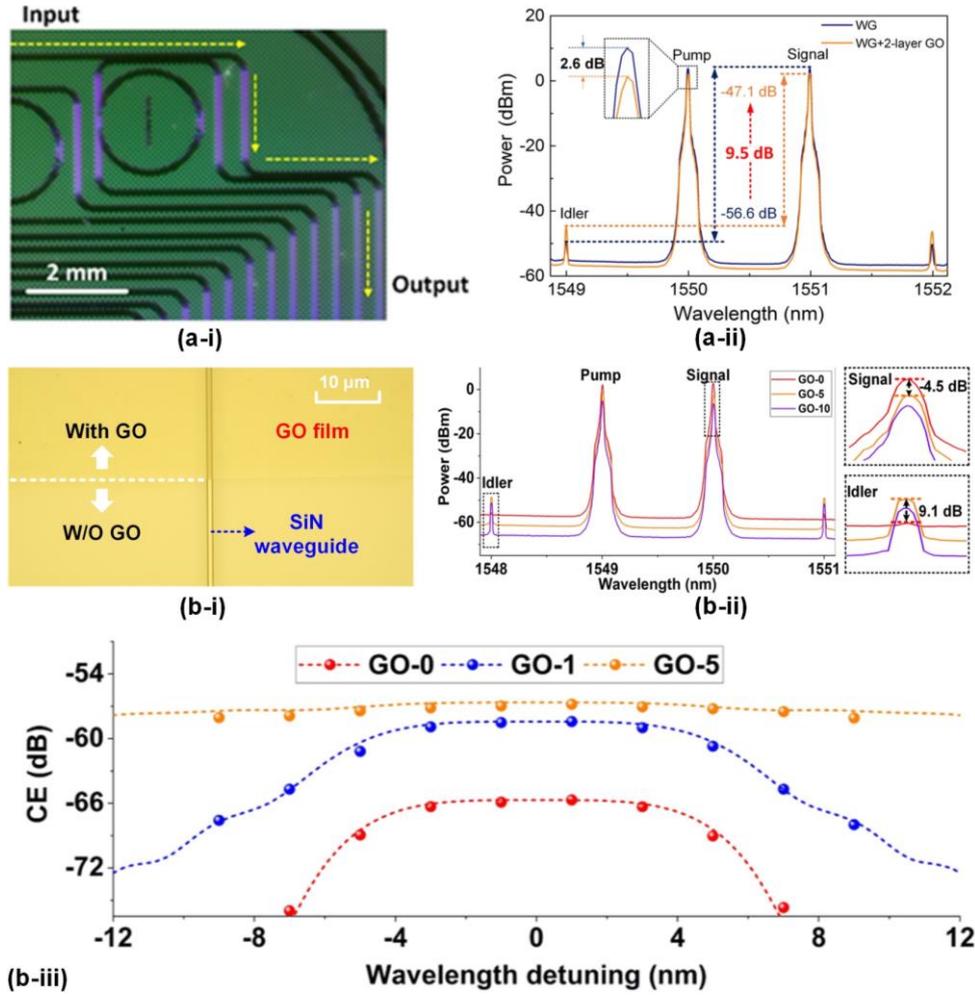

Figure 13. Enhanced FWM in GO-coated integrated waveguides. (a) FWM in GO-coated doped silica waveguides [56]: (i) microscopic image of the waveguides uniformly coated with 2 layers of GO; (ii) FWM spectra for the uncoated and coated waveguides. (b) FWM in GO-coated SiN waveguides [78]: (i) microscopic image of a SiN waveguide with 10 layers of patterned GO; (ii) FWM spectra for the bare waveguide and the waveguides with 5 and 10 layers patterned GO films; (iii) measured (data points) and fit (dashed curves) CE versus wavelength detuning for the bare SiN waveguide (GO-0), the uniformly coated device with 1 layer of GO (GO-1), and the patterned device with 5 layers of GO (GO-5).

Layered 2D GO films have also been integrated with MRRs (**Fig. 14(a)**) to achieve further enhancement in FWM efficiency [54]. Compared with waveguides, FWM in MRRs provides significantly enhanced CE due to the resonant enhancement of the optical fields [13, 153]. **Figs. 14(b-i)** and **(b-ii)** show the FWM spectra for doped silica MRRs uniformly coated with 1−5 layers of GO and patterned with 10−50 layers of GO, respectively. An enhancement in the FWM CE, relative to the uncoated device, of ~7.6-dB for a uniformly coated device with 1 layer of GO and ~10.3-dB for a patterned device with 50 layers of GO was achieved. **Figs. 14(c-i)** and **(c-ii)** show the FWM spectra versus pump-signal wavelength detuning (Δλ) for these two devices, showing only a slight decrease in CE of < 1.6 dB when Δλ equals to 20 free spectral ranges (FSRs). This reflects the low dispersion of the GO-coated doped silica MRRs, which enables effective phase matching for broadband FWM.

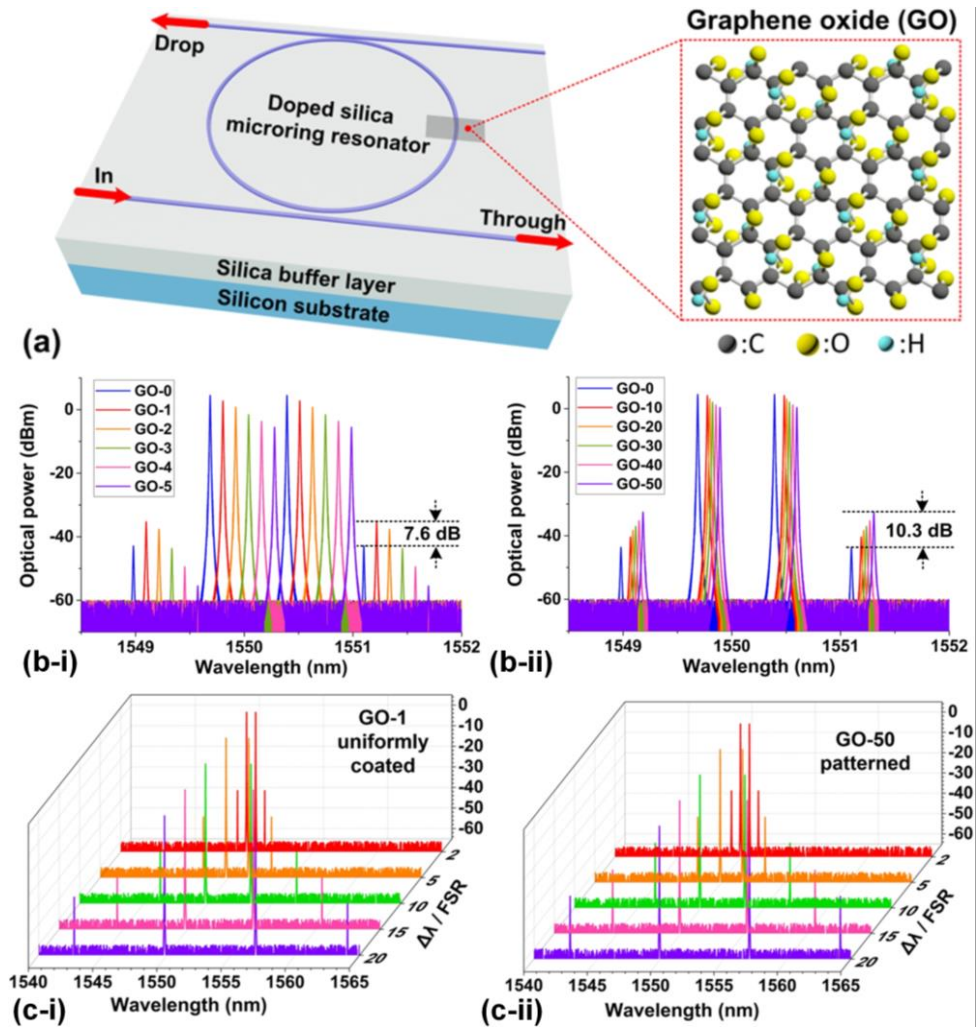

Figure 14. Enhanced FWM in GO-coated doped silica MRRs [54]. (a) Device schematic. (b) FWM spectra at a pump power of 22 dBm for the MRRs with (i) 1−5 layers of uniformly coated and (ii) 10−50 layers of patterned GO films. (c) FWM spectra versus pump-signal wavelength detuning (Δλ) for the MRRs (i) uniformly coated with 1 layer of GO and (ii) patterned with 50 layers of GO.

SPM is another important Kerr nonlinear optical effect that occurs when an ultrashort high-peak-power optical pulse propagates through a nonlinear medium. The self-induced change in refractive index from the Kerr effect produces a phase shift in the pulse and hence modifies the pulse's spectrum. It has wide applications in broadband optical sources, spectroscopy, bio-imaging, pulse compression, and optical coherence tomography [154].

SOI nanowires have been reported with integrated GO films to demonstrate enhanced SPM [79]. **Fig. 15(a)** shows a schematic of an SOI nanowire conformally coated with a GO film. A microscopic image of the GO-coated SOI chip is shown in **Fig. 15(b)**. Windows were opened on the silica upper cladding to enable film coating. Two GO patterns were fabricated – (i) 2.2-mm-long with 1−3 layers and (ii) 0.4-mm-long with 5−20 layers. The length of the SOI nanowires was 3 mm. **Figs. 15(c)** and **(d)** show the SPM results measured using a pulsed fiber laser. The GO-coated SOI nanowires showed much more significantly broadened spectra as compared with the bare SOI nanowire, achieving high broadening factors (BFs) [155] (3.75 for 2 layers of GO and 4.34 for 10 layers of GO) and reflecting a significantly increased Kerr nonlinearity.

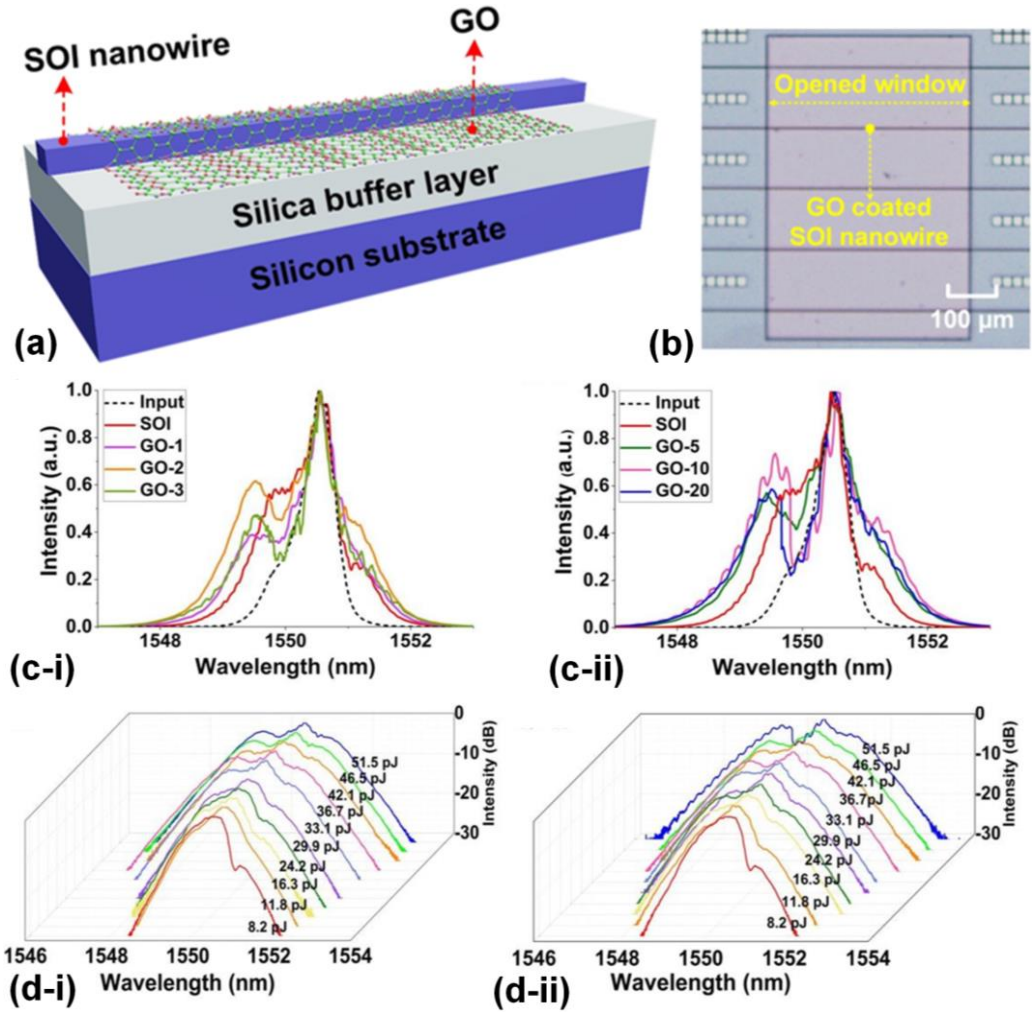

Figure 15. Enhanced SPM in GO-coated SOI nanowires [79]. (a) Device schematic. (b) Microscope image of a GO-coated SOI chip. (c) Normalized spectra of optical pulses before and after going through the bare and hybrid waveguides at a coupled pulse energy of ~51.5 pJ. (d) Optical spectra measured at different pulse energies. In (c) and (d), (i) and (ii) show the results for the hybrid waveguides with 2.2-mm-long, 1−3 layers of GO and with 0.4-mm-long, 5−20 layers of GO, respectively.

**Table 2** compares the performance of doped silica, SiN, and silicon waveguides, where the dimensions were quite different, incorporated with GO films. The silicon waveguide had the smallest waveguide dimensions as well as the tightest mode confinement and the highest index contrast, resulting in significantly increased mode overlap with the GO film. This resulted in a significantly increased nonlinear parameter $\gamma$, but also the largest excess propagation loss induced by the GO film. Mode overlap is a key factor for optimizing the trade-off between the Kerr nonlinearity and loss when introducing 2D layered GO films onto different integrated platforms to enhance the nonlinear optical performance.

**Table 2. Performance comparison of doped silica, SiN, and silicon waveguides integrated with 2D layered GO films. WG: waveguide. FOM: figure of merit.**

|  | Silicon-GO hybrid WG [79] | SiN-GO hybrid WG [78] | Doped silica-GO Hybrid WG [56] |
|---|---|---|---|
| $n_0$ of bare WG | 3.45 | 1.99 | 1.66 |
| $n_2$ of bare WG ($m^2$/W) | $6.03 \times 10^{-18}$ | $2.61 \times 10^{-19}$ | $1.28 \times 10^{-19}$ |
| Nonlinear FOM of bare WG [a] | 0.3 | >>1 | >>1 |
| Waveguide dimension (μm) | Width: 0.5 Height: 0.22 | Width: 1.60 Height: 0.66 | Width: 2.00 Height: 1.50 |
| Propagation loss of bare WG (dB/cm) | 4.3 | 3.0 | 0.24 |
| $EPL_{GO-1}$ [b] (dB/cm) | 20.5 | 3.1 | 1.0 |
| $\gamma_{WG}$ [c] ($W^{-1}m^{-1}$) | 288 | 1.51 | 0.28 |
| $\gamma_{hybrid}$ [d] ($W^{-1}m^{-1}$) | GO-1: 668 GO-10: 2905 | GO-1: 13.14 GO-10: 167.14 | GO-1: 0.61 GO-10: n/a [e] |
| $n_2$ of GO ($m^2$/W) [f] | $1.22 \times 10^{-14}$ ~$1.42 \times 10^{-14}$ | $1.28 \times 10^{-14}$ ~$1.41 \times 10^{-14}$ | $1.5 \times 10^{-14}$ |
| Effective nonlinear FOM of hybrid WG | GO-20: 5.7 [g] | >>1 | >>1 |

a) The nonlinear FOM is defined as FOM = $n_2 / (\lambda \cdot \beta_{TPA})$, where $\lambda$ is the wavelength and $\beta_{TPA}$ is the TPA coefficient [14].
b) $EPL_{GO-1}$: excess propagation loss induced by GO for the hybrid waveguide with 1 layer of GO.
c) $\gamma_{WG}$: nonlinear parameter of the bare waveguide.
d) $\gamma_{hybrid}$: nonlinear parameter of the hybrid waveguide with 1 layer (GO-1) and 10 (GO-10) layers of GO.
e) Only hybrid waveguides with 1-5 layers of GO were characterized.
f) The $n_2$ values of GO are extracted from the nonlinear FWM or SPM experiments.
g) GO-20: the result for the hybrid waveguide with 20 layers of GO.

In contrast to the real part of the third-order optical susceptibility (i.e., Re($\chi^{(3)}$)), which is responsible for Kerr nonlinear processes, the imaginary part (i.e., Im($\chi^{(3)}$)) accounts for nonlinear absorption such as multiphoton absorption and potentially also SA [62], although generally SA arises from real photo-generated carrier effects and not virtual processes. In the past decade, SA of 2D materials has been widely exploited for passively mode-locked fiber lasers with broad applications from industrial processing to fundamental research [62, 63]. Although GO has shown relatively weak SA compared to graphene, many GO-based passively mode-locked fiber lasers have been realized by using the solvent dispersibility and chemical reduction of GO [156-159]. Implementing SA-based devices in integrated platforms is a powerful method to achieve critical nonlinear elements for advanced photonic integrated circuits including integrated mode-locked lasers [160], broadband ultrafast all-optical modulators [161], and photonic neural networks [162].

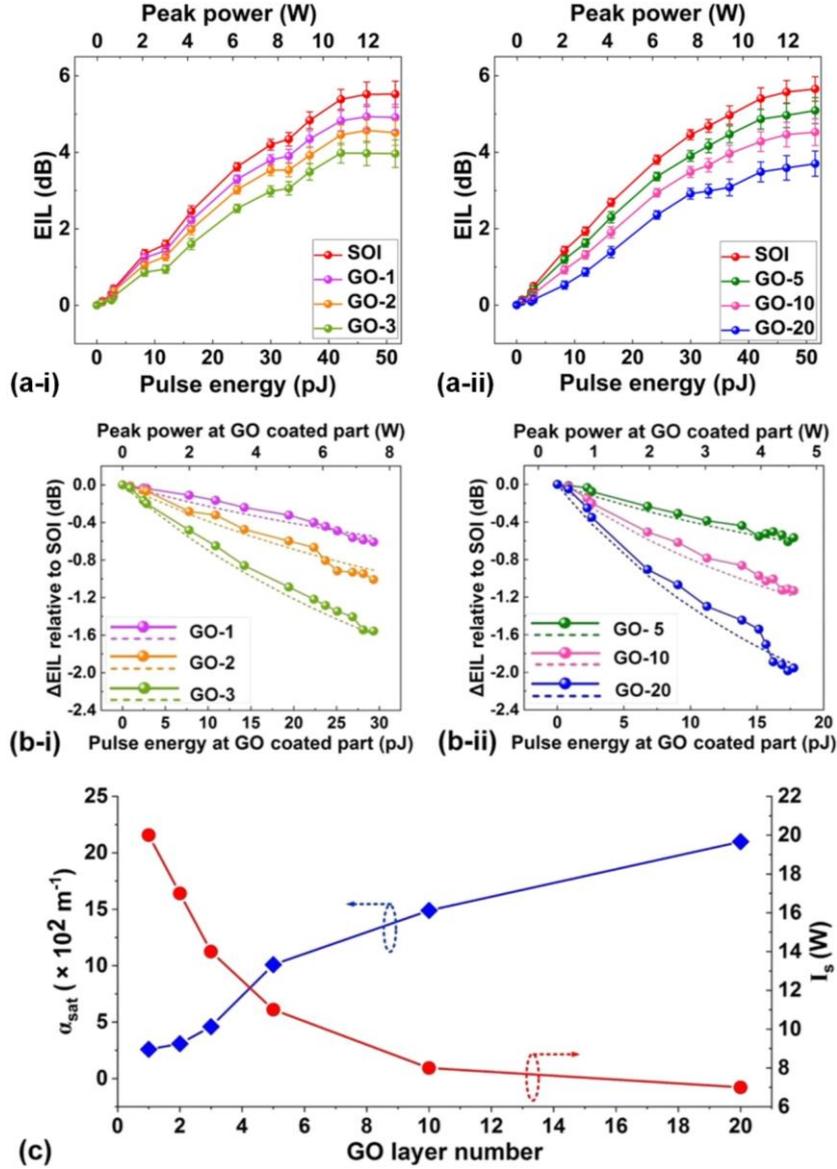

Figure 16. SA in GO-coated SOI nanowires [79]. (a) Power-dependent excess insertion loss (EIL) of optical pulses after transmission through the bare and hybrid SOI nanowires. (b) Measured (solid curves) and fit (dashed curves) ΔEIL relative to bare SOI nanowires versus pulse energy incident. In (a) and (b), (i) shows the results for the waveguides with 2.2-mm-long, 1−3 layers of GO and (ii) shows the results for the waveguides with 0.4-mm-long, 5−20 layers of GO. (c) Fit $α_{sat}$ and $I_s$ versus GO layer number.

Recently, Zhang *et.al* investigated SA in GO-coated SOI nanowires [79], measuring the power-dependent excess insertion loss relative to the bare SOI nanowires (ΔEIL) as a function of the coupled pulse energy. While the overall insertion loss increases (**Fig. 16(a)**) due to TPA and free carrier absorption in silicon [12, 163], ΔEIL decreases with pulse energy (**Fig.**

**16(b)**) due to the influence of the nonlinear loss in the GO films – a trend consistent with SA. The fit curves were based on:

$$\alpha_{NL\text{-}GO} = \alpha_{sat}/(1 + \frac{|A|^2}{I_s}) \qquad (1)$$

where $\alpha_{sat}$ is the SA coefficient, $I_s$ is the saturation intensity, and $A$ is the slowly varying temporal envelope of the optical pulse. **Fig. 16(c)** shows that $\alpha_{sat}$ increases with GO layer number, whereas $I_s$ shows the opposite trend, reflecting the dependence of SA in GO films with layer number, where the SA increases and the power threshold decreases for thicker films.

*4.6 Light emitting devices*

For light emitting devices, the heterogeneous atomic and electronic structure of GO results in a very broadband PL in the near-infrared, visible and ultraviolet wavelength regions [19, 26, 39]. Benefitting from its large and direct material bandgap, the PL in GO is also much stronger than that of graphene.

In 2012, Lee *et al.* [164] demonstrated enhanced performance for polymer light-emitting diodes (LEDs) with GO interlayers (**Fig. 17(a)**), achieving a 220% increase in luminous efficiency and a 280% increase in PCE compared to the devices without GO. The GO interlayers prevented significant quenching of the radiative excitons between the emissive polymer and the GO layer and maximized hole-electron recombination within the emissive layer, ultimately leading to significantly improved device performance.

Han *et al.*[165] subsequently demonstrated that embedded rGO in a gallium nitride (GaN) LED alleviated the self-heating issues due to its high thermal conductivity (**Fig. 17(b)**. Scalable rGO microscale pattern was generated by using a combination of facile lithography and spray-coating methods, which acted as a buffer layer for the epitaxial lateral overgrowth of high-quality GaN, thus offering excellent heat dissipation while maintaining electrical and optical properties superior to that of its conventional counterpart.

In 2015, Wang *et al.* [166] demonstrated spectrally tunable luminescence from the blue (~450 nm) to red (~750 nm) in GO/rGO-based field-effect LEDs. **Fig. 17(c)** shows the fabricated device as well as the gate-tunable luminescence spectra. The device formed on a laser-scribed GO surface consists of a series of rGO nanoclusters with different sizes, which could selectively stimulate a single-color luminescence by controlling the doping level via gate modulation. A high brightness of up to 6,000 cd m$^{-2}$ was achieved, with an efficiency of about 1 %.

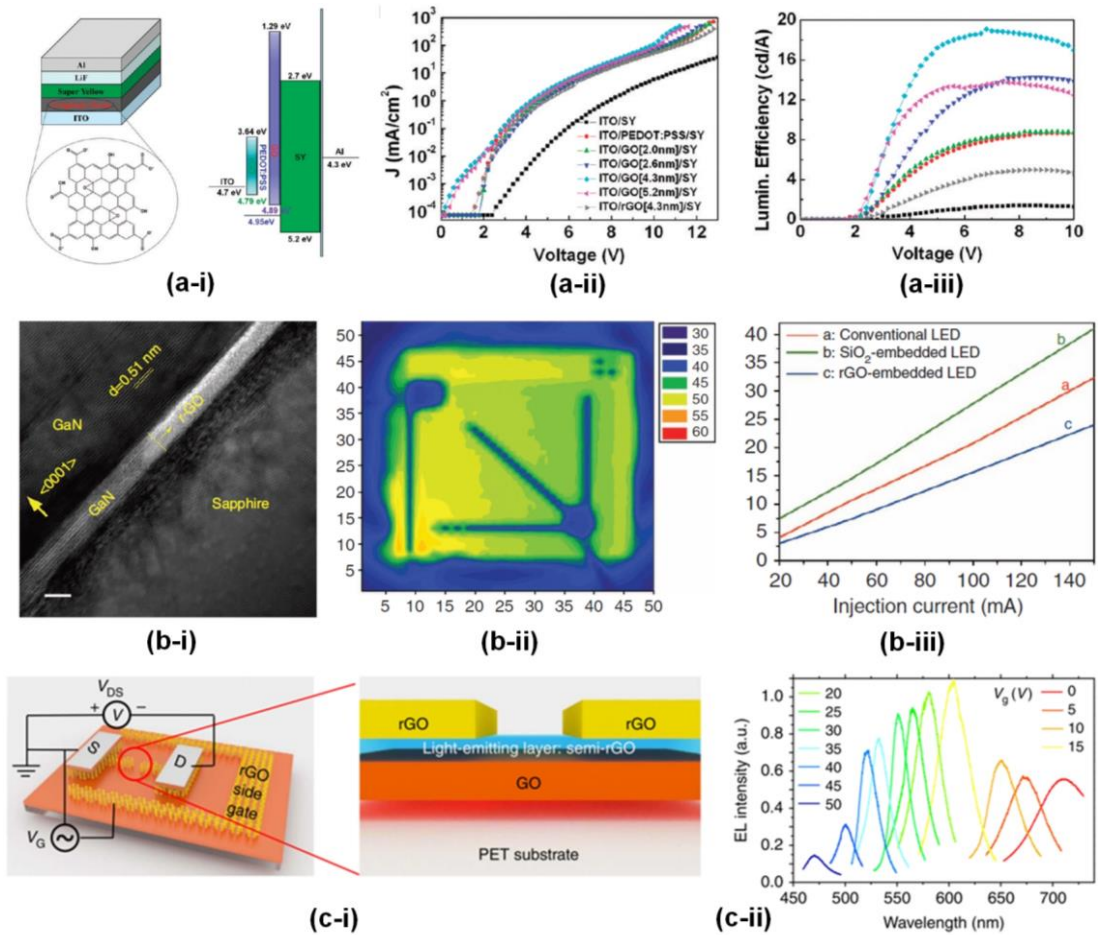

Figure 17. GO-based light emitting diodes (LEDs). (a) A polymer LED with a GO film as a hole transport interlayer [164]: (i) device schematics; (ii) – (iii) current density and luminance efficiency versus applied voltage for various charge transport layers, respectively. (b) A rGO-embedded gallium nitride (GaN) LED [165]: (i) TEM image of the GaN/rGO/sapphire interface; (ii) temperature distribution on the surface of rGO-embedded chip; (iii) Variations in junction temperature as a function of injection current. (c) A GO/rGO-based field-effect LED [166]: (i) device schematic; (ii) typical luminescence spectra for varied gate bias from 0 to 50 V.

*4.7 Photodetectors (PDs)*

High-performance broadband PDs are critical for a variety of applications, including optical communication, imaging, remote sensing, environmental monitoring, astronomical detection, photometers, etc [6]. In the past decade, graphene-based PDs have attracted strong interest due to their exceptional optoelectronic properties, including broadband ultrafast response, strong electron–electron interaction, and photocarrier multiplication [8, 167-170]. Since pristine GO is a dielectric with low electronic conductivity and light absorption, it cannot be directly used for photo detection. Nevertheless, the reduction of GO is a more scalable and cost-effective approach for mass production of graphene-like materials as compared with the traditional mechanical and vacuum-based preparation methods, enabling many PDs based on rGO [171-184]. Moreover, rGO also has its own advantages such as high flexibility in engineering its material properties and great compatibility with a variety of substrates.

**Table 3. Performance comparison of rGO PDs. EQE: external quantum efficiency.**

| Wavelength (nm) | Maximum photo responsivity (mA/W) | Maximum EQE | Bias Voltage (V) | Response / recover time (s) | Ref. |
|---|---|---|---|---|---|
| 370 | 860 | -- [b)] | 10 | -- / -- | [171] |
| 532 | -- | -- | 8 | $2.5 \times 10^{-4}$ / $5.0 \times 10^{-4}$ | [173] |
| 808 | -- | -- | ±4 | 2.36 / 2.54 | [172] |
| 1550 | 4 | 0.3% | 2 | 2 / 23 | [175] |
| 360 | 120 | 40% | 1 | 1800 / 2100 | [174] |
| 895 | 700 | 97% | 19 | 2 / 6 | [176] |
| 850 | 0.0043 | -- | 1 | 70 / 93 | [177] |
| 532 – 119000 | 9 | -- | 1 | 10 – 27 / 3 – 26 | [178] |
| 445 | 128.72 | 0.31% | -0.5 | $2.6 \times 10^{-4}$ / $5.4 \times 10^{-4}$ | [179] |
| 633 | 0.27 | 0.05% | 0 | 42.5 / 56.4 | [182] |
| 600 | 1520 | -- | 0 | 0.002 / 0.0037 | [181] |
| 370 | $3.1 \times 10^7$ | $1.04 \times 10^7$ % | 5 | 2.13 / 3.44 | [180] |
| 375 – 1064 | 428 | 100% | 1 | 0.61 – 0.99 / -- | [184] |
| 375 – 1064 | 51.46 | -- | 0.5 | 0.04 – 0.146 / 0.066 – 0.139 | [183] |

a) --: The results are not provided.

**Table 3** compares the performance of the state-of-the art rGO PDs. In 2010, Ghosh *et al.* [172] demonstrated an infrared PD based on large-area rGO sheets with a laser spot position dependent photo response (**Fig. 18(a)**). By engineering the defects in rGO, Chang *et al.* [176] subsequently achieved significantly improved photo responsivity (700 mA/W, over one order of magnitude higher than that of pristine graphene) and external quantum efficiency (97%) in a rGO photo transistor (**Fig. 18(b)**). In 2013, Cao *et al.* [177] characterized the photo responsivity and response time of rGO PDs as a function of layer number (**Fig. 18(c)**). In 2014, a PD based on self-assembled rGO-silicon nanowire array heterojunctions was demonstrated (**Fig. 18(d)**) [178], operating in an ultrabroad band from the visible (532 nm) to terahertz region (2.52 THz, or a wavelength of 118.8 μm) at room temperature. In 2016, Li *et al.* [181] demonstrated a self-powered PD consisting a p-n vertical heterojunction between a drop-casted rGO thin film and n-doped silicon (**Fig. 18(e)**), which allowed efficient transfer of photogenerated charge carriers and ultimately resulted in high photo responsivity (1520 mA/W) and fast response time (on the order of $10^{-3}$ s). In 2017, Gan *et al.* [183] demonstrated fully suspended rGO PDs with different annealing temperatures (**Fig. 18(f)**), achieving time responses that are 1−4 orders of magnitude faster than comparable rGO PDs supported by substrates.

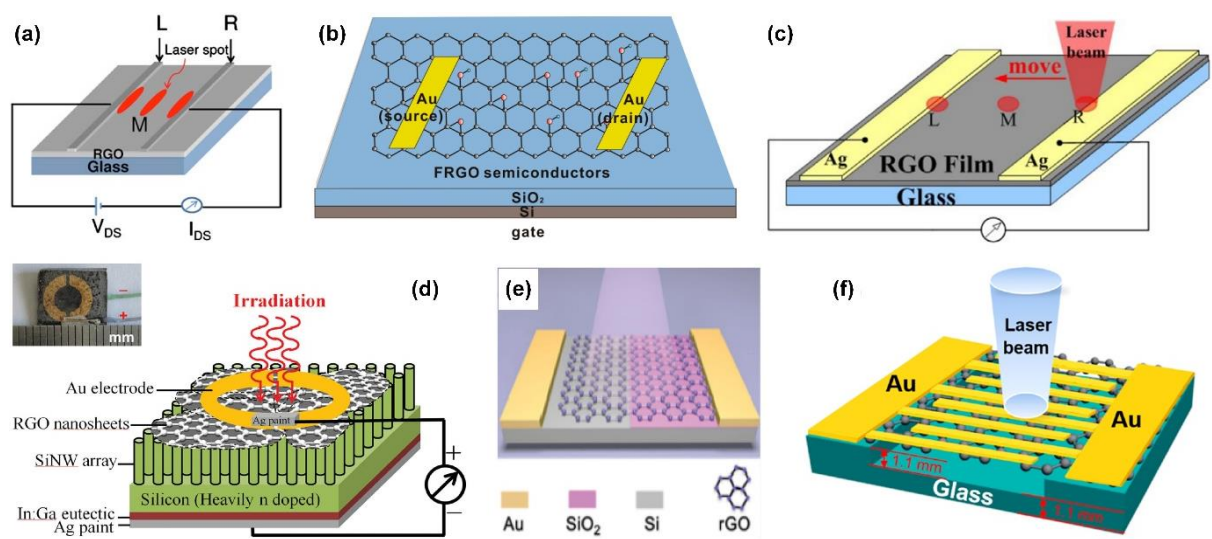

Figure 18. rGO photodetectors (PDs). (a) A position sensitive PD based on large-area rGO sheets [172]. (b) A photo transistor based on few-layer rGO with controlled defects [176]. (c) A near-infrared PD based on a thin rGO film with controlled number of bilayers [177]. (d) An ultra-broadband PD based on self-assembled rGO-silicon nanowire array heterojunctions [178]. (e) A self-powered PD based on rGO/n-silicon vertical heterojunction [181]. (f) A PD based on a fully suspended rGO thin film [183].

## 5. Challenges and perspectives

The past decade has witnessed the rapid development of integrated GO photonics, including investigation of the material properties as well as demonstrating advanced hybrid integrated devices. Though the distinctive optical properties and strong capability for on-chip integration of GO films have already enabled many functional hybrid integrated photonic devices with new capabilities, there is still much room for development in material properties, device fabrication, and in creating new applications.

In terms of material properties, although the propagation loss of GO hybrid waveguides is now about two orders of magnitude lower than comparable devices integrated with graphene, it is still not enough for some nonlinear applications such as optical micro-comb generation [185]. While in theory GO films with a bandgap > 2 eV should have negligible absorption at near-infrared wavelengths, the reality is that in practice the absorption of GO films is still significant, mainly induced by light absorption from localized defects as well as scattering loss due to film unevenness and imperfect contact between the different layers. Unlike graphene, however, this is not a fundamental property and so can be reduced by optimizing the GO synthesis and coating processes, for example by using GO solutions with reduced flake sizes and increased purity. Reducing the loss of GO films would not only improve the performance of FWM and SPM, but also potentially the parametric gain needed for on-chip optical micro-comb generation as well as SA for subsequent micro-comb mode locking [185-207].

Although a substantial body of work has been carried out to investigate the nonlinear optical response of GO, much of this work, particularly that centered on the Kerr nonlinearity or nonlinear absorption, has been semi-empirical in nature with little in-depth study of the underlying physics. Many physical insights regarding the linear, and particularly the nonlinear, optical properties of GO or rGO remain unexplored, such as the anisotropic optical nonlinearity and the interplay between the Kerr nonlinearity and nonlinear absorption, hinting at much exciting research to come. There is a synergy between the on-chip integration of GO and the investigation of its material properties. Integrated platforms with high fabrication yield and production at scale provide powerful and mature devices with which to investigate the material properties of 2D GO films, while this is challenging for Z-scan measurements because of the weak response of ultrathin 2D films [65, 77]. On the other hand, a deep understanding of the material properties of GO will allow the full exploitation of its significant potential for integrated photonic devices.

With respect to device fabrication, although the conformal coating of GO films on single-mode silicon nanowire waveguides (with a cross section of 500 nm × 220 nm) has been achieved [79], it is still difficult to achieve conformal coating in structures with feature sizes < 100 nm such as slot waveguides. This is mainly limited by the GO flake size used in solution-based self-assembly, although this can be reduced via oxidation and vigorous ultrasonics [21]. Slot waveguides, which are capable of strongly confining light within subwavelength slots, provide an ideal device structure to introduce many materials onto integrated platforms [137, 208]. The slot structure could significantly enhance the field intensity and mode overlap for light-matter interaction [56], enabling better exploitation of GO's superior material properties such as its Kerr nonlinearity and SA.

A key issue for fabricating integrated devices is the patterning resolution, and there are several factors contributing to this for GO films, including the film thickness, lithography resolution, size of the GO flakes, and thickness of the photoresist. For thin GO films (< 10 layers), pattern resolution is mainly limited by lithography as well as the GO flake size, whereas for thick films (> 50 layers), the thickness itself becomes the dominant factor, leading to a trade-off between patterning resolution and film thickness. As compared with photolithography, electron-beam lithography has a higher resolution, although at the expense of longer required exposure times. E-beam lithography has been used to write patterns on 300-nm-thick photoresist, creating short pattern lengths of ~150 nm and ~500 nm for 2 and 30 layers of GO, respectively [53].

For the reduction of GO films, ideally, it could remove the OFGs and restore the $sp^2$ carbon network like that of graphene. However, practically defects form in removing the OFGs, introducing a difference in the properties between rGO and graphene. In fact, the defects can form during chemical oxidation even before reduction, and so it is extremely challenging to obtain high purity graphene strictly through GO reduction, although much work has been devoted to achieving this [20]. Despite this, reduction of GO films yields significant flexibility in tailoring device properties [41, 42, 44, 48, 166], which is difficult to achieve for integrated photonic devices. Finer control of GO reduction is also needed, particularly for laser reduction of ultrathin 2D films. Further, in addition to GO reduction via the removal of OFGs, heteroatom doping in GO via special treatments in a dopant environment [209], which has been used in fabricating GO-based electronic devices [35] but has not yet been widely adopted for photonic devices, could also lead to the engineering of GO's optical properties for many new applications.

There are many applications of hybrid GO integrated photonic devices beyond what is discussed in Section 4, such as patterning or reducing GO films to engineer the waveguide dispersion for broadband phase matching, that, together with its high Kerr nonlinearity, could yield very efficient super-continuum generation. In contrast to conventional phase matching based on anomalous dispersion, needed for materials with a positive $n_2$, the negative $n_2$ of rGO offers the possibility to achieve phase matching in waveguides with normal dispersion. This is particularly useful for < 400-nm-thick SiN waveguides with normal dispersion [210], where achieving a negative net $n_2$ would render them more useful for nonlinear optics. Moreover, the SA of GO can also be used for broadband ultrafast all-optical modulators [161, 211]. Along with advances in understanding and controlling of GO's optical properties as well as improvement in fabrication and reduction processes, we believe that GO hybrid integrated photonic devices will prove to be extremely attractive for many existing and new applications in cross-disciplinary fields, ultimately bridging the gap from laboratory research to practical industrial applications.

## 6. Conclusion

Integrated GO photonics represents a nascent and promising field at the intersection of integrated optics and GO material science. This active and fast-growing field, with its roots in the superior optical properties of GO and mature integrated device platforms, has experienced significant advances over the past decade, particularly in the large-scale, highly precise on-chip integration of GO together with the flexible manipulation of its material properties. In this review, we have discussed the optical properties of GO, summarizing the approaches to on-chip integration of GO films. We review the diverse applications of GO hybrid integrated photonic devices, including both passive (linear and nonlinear) and active devices. We also discuss the strong potential as well as the challenges that remain in this field. The on-chip integration of GO is a harbinger of a new generation of compact, manufacturable hybrid integrated photonic devices, offering high performance and enormous new possibilities for scientific research and industrial applications.

## Acknowledgement

This work was supported by the Australian Research Council Discovery Projects Programs (No. DP150102972 and DP190103186), the Swinburne ECR-SUPRA program, the Industrial Transformation Training Centers scheme (Grant No. IC180100005), and the Beijing Natural Science Foundation (No. Z180007).